\documentclass[journal]{IEEEtran}
\usepackage{amsmath}
\usepackage{amssymb}
\usepackage{amsfonts}
\usepackage{mathtools}
\usepackage{graphicx}
\usepackage{epsfig}
\usepackage{psfrag}
\usepackage{tabulary}
\usepackage{booktabs}
\usepackage[skip=0pt]{caption}
\usepackage{pifont}
\usepackage{color}
\usepackage{bm}
\usepackage{subcaption}
\usepackage{lettrine}
\usepackage{url}
\usepackage[colorlinks,linkcolor=blue,anchorcolor=blue,citecolor=blue,urlcolor=black,breaklinks]{hyperref}
\usepackage{breakurl}
\usepackage{cite}
\usepackage{epstopdf}
\newcommand{\mv}[1]{\mbox{\boldmath{$ #1 $}}}

\def\({\left(}
\def\){\right)}

\def\b0{{\mathbf{0}}}

\def\mP{{\mathbb{P}}}

\def\mR{{\mathbb{R}}}

\def\cF{\mathcal{F}}
\def\cG{\mathcal{G}}

\def\cK{\mathcal{K}}
\def\cL{\mathcal{L}}

\def\cN{\mathcal{N}}

\def\cR{\mathcal{R}}

\def\cU{\mathcal{U}}

\def\cW{\mathcal{W}}

\newcommand{\inm}{\in\mathcal}

\newlength{\arrow}
\newcolumntype{L}[1]{>{\raggedright\let\newline\\\arraybackslash\hspace{0pt}}m{#1}}
\newcolumntype{C}[1]{>{\centering\let\newline\\\arraybackslash\hspace{0pt}}m{#1}}
\newcolumntype{R}[1]{>{\raggedleft\let\newline\\\arraybackslash\hspace{0pt}}m{#1}}
\usepackage{ragged2e}

\newcommand{\cacheprob}{c_{k,f}}
\newcommand{\cacheprobcomp}{\bar{c}_{k,f}}
\newcommand{\reqprob}{r_{k,f}}

\title{Cooperative Local Caching Under Heterogeneous File Preferences}
\author{Yinghao Guo, Lingjie Duan,~\IEEEmembership{Member,~IEEE} and Rui Zhang,~\IEEEmembership{Senior Member,~IEEE}
\thanks{Y. Guo is with the Department of Electrical and Computer Engineering, National University of Singapore (e-mail: yinghao.guo@u.nus.edu).}
\thanks{L. Duan is with the Engineering Systems and Design Pillar, Singapore University of Technology and Design (e-mail: lingjie\_duan@sutd.edu.sg).}
\thanks{R. Zhang is with the Department of Electrical and Computer Engineering, National University of Singapore (e-mail:elezhang@nus.edu.sg). He is also with the Institute for Infocomm Research, A*STAR, Singapore.}}

\begin{document}

\maketitle\thispagestyle{empty}
\newtheorem{definition}{\underline{Definition}}[section]
\newtheorem{fact}{Fact}
\newtheorem{assumption}{Assumption}
\newtheorem{theorem}{\underline{Theorem}}[section]
\newtheorem{lemma}{\underline{Lemma}}[section]
\newtheorem{corollary}{\underline{Corollary}}[section]
\newtheorem{proposition}{\underline{Proposition}}[section]
\newtheorem{example}{\underline{Example}}[section]
\newtheorem{remark}{\underline{Remark}}[section]
\newtheorem{algorithm}{\underline{Algorithm}}[section]
\newtheorem{aspect}{\underline{Aspect}}

\begin{abstract}
  Local caching is an effective scheme for leveraging  the memory of the mobile terminal (MT) and short range communications to save the bandwidth usage and reduce the download delay in the cellular communication system. Specifically, the MTs first cache  in their local memories in off-peak hours and then exchange the requested files with each other in the vicinity during peak hours. However, prior works largely overlook MTs' heterogeneity in file preferences and their selfish behaviours. In this paper, we practically categorize the MTs into different interest groups according to the MTs' preferences. Each group of MTs aims to increase the probability of successful file discovery from the neighbouring MTs (from the same or different groups). Hence, we define the groups' utilities as the probability of successfully discovering the file in the neighbouring MTs, which should be maximized by deciding the caching strategies of different groups. By modelling MTs' mobilities as homogeneous Poisson point processes (HPPPs), we analytically characterize MTs' utilities in closed-form. We first consider the fully cooperative case where a centralizer helps all groups to make caching decisions. We formulate the problem as a weighted-sum utility maximization problem, through which the maximum utility trade-offs of different groups are characterized. Next, we study two benchmark cases under selfish caching, namely, partial and no cooperation, with and without inter-group file sharing, respectively. The  optimal caching distributions for these two cases are derived.  Finally, numerical examples are presented to compare the utilities under different cases  and show the effectiveness of the fully cooperative local caching  compared to the two benchmark cases.
\end{abstract}

\section{Introduction}\label{sec:intro}
{\lettrine{I}{t} is estimated that the global mobile data traffic will exceed 24.3 exabytes per month by 2019\cite{CISCOVNI}.  Moreover, the data traffic is increasingly  concentrated to hotspots and wireless networks are getting more  congested, especially during peak hours \cite{informa2008mobile}. Although there are various techniques to increase the network capacity, such as the densification of network, millimeter wave communications and massive MIMO, they do not fully exploit the characteristics of the mobile data traffic: wireless data traffic fluctuates significantly during a day. Also, the mobile video accounts for about 70\% of the total data traffic and a large amount of video traffic is due to the downloading of very few popular files\cite{CISCOVNI}. As a solution, {\it local caching} \cite{LocalCachingComMag} has been proposed recently as an effective approach for addressing the throughput bottleneck of wireless networks. It trades off the precious bandwidth resource or the backhaul links with the data storage at the network's edge devices (e.g., smartphones), which is usually highly under-utilized and the price of which is getting lower. Popular files, which are usually a small amount of files compared to the size of the whole library,  can be pro-actively downloaded in the edge devices during off-peak hours and the edge devices share the cached files locally with each other during  the peak hours.  In this way, local caching alleviates the traffic congestion during the peak hours, hence flattening the data traffic and enhancing the robustness of the network.

Traditional caching has long been proposed for reducing the downloading delay in the wired network\cite{Dowdy}. It exploits the opportunity that the requested file may reside in the local cache and the file can be directly obtained without resorting to the remote server. While for cooperative local caching, the main difference is that its performance also depends on the aggregate cache of all the edge devices of the local network due to their cooperation \cite{FundLimitofCaching}. However, from the work on cooperative local caching in the literature, we notice that prior works largely assume that different MTs have the same preference over the files  and they cooperate with each other without self-interests. Motivated by this, in this paper, we study the cooperative local caching under heterogeneous file preferences, where different MTs have different preferences over the files in general. Given such heterogeneity, some selfish MTs may want to cache only according to their own interests without considering the other MTs. We will study the impacts of such selfish behaviours on the performances of both themselves and the other MTs.}

The main contributions of this paper are summarized as follows.
\begin{itemize}
\item{\emph{Practical modelling of heterogeneous file preferences:}} In Section \ref{sec:system_model}, we practically model the MTs' heterogeneous file preferences and categorize them into different interest groups according to their request preferences over the files. We also model their spatial locations as homogeneous Poisson point processes (HPPPs) with different spatial densities. Under the above set-up, closed-form expression of each group's utility, defined as the probability of the file discovery in the local or neighbouring cache, is derived. The feasible utility region, which constitutes all the achievable utility trade-offs among different groups, is then defined.

\item{\emph{Full cooperation under centralized coordination:}}
In Section \ref{sec:FullCoop}, when all MT groups are fully cooperative, we study the optimal caching distribution decided by the centralizer. In order to characterize the complete feasible utility region for different groups, we formulate the problem as a weighted-sum utility maximization problem. Algorithms based on coordinate descent method are proposed for solving the problems for the groups with positive and zero weights, respectively.

\item{\emph{Benchmark cases under partial and no cooperation:}} In Section \ref{sec:PartialCoopNoCoop}, comparing with full cooperation, we study two benchmark cases under selfish caching, namely, partial and no cooperation.  Optimal caching solutions in different cases are obtained by utilizing their Karush-Kuhn-Tucker (KKT) conditions.

\item{\emph{Performance comparison via simulations:}} In Section \ref{sec:numerical}, we provide numerical results to validate the convergence of the coordinate descent algorithm and compare the utilities under different system parameters.  By obtaining the complete feasible utility region, we show a significant utility increase in the scheme of full cooperation as compared with partial and no cooperation.  We also perform a network simulation where the real-time operations of MTs are simulated under the evolution of the system. The effectiveness of our proposed scheme and the accuracy of the modelling are shown.
\end{itemize}

\subsection{Related Work}
{For cooperative local caching, depending on which part of network edge to cache the popular files, the literature can be categorized into two lines of works, namely caching on the base station (BS) side  \cite{TITFemto,AdaptiveVIdeoStreaming,TaoMeixiaYuWei} and caching on  the MT side \cite{TITD2Dcaching,JiCM13,JiCM14DundLimitD2D,Multihop}.
First, for local caching on the BS side, popular files are selected and cached in the BSs and the BSs cooperatively serve the MTs in the cellular network.
\cite{TITFemto} considers that BSs cooperatively serve MTs by deterministically caching files under the coded and uncoded scenarios.  The drawback of this approach is that the location information of the MTs is assumed and the cache needs to be updated accordingly.
In \cite{AdaptiveVIdeoStreaming}, an efficient caching protocol is proposed, which allows the MTs to dynamically select BSs that can serve  the MTs and adaptively adjust the quality of transmission.
\cite{TaoMeixiaYuWei} considers a broadcasting scheme where multiple BSs cooperatively serve MTs with sparse multicast beamforming for reducing the backhaul data transmission in the cloud radio access network.

Next, for local caching on the MT side or device-to-device (D2D) local caching,  files are pre-cached in the MTs during the off-peak hours and MTs share files with each other in the file delivery during peak hours.
{\cite{TITD2Dcaching} derives the power scaling law of the network capacity with respect to the range of the communications.  Yet, \cite{TITD2Dcaching} only discusses the metric of throughput, where all users are required to be served for any request. In light of this, \cite{JiCM13} considers the file sharing outage in the local caching and the fairness issue of MTs by characterizing the capacity-outage trade-off.} \cite{JiCM14DundLimitD2D} extends the work on the BS side caching in \cite{FundLimitofCaching} to the MT side by proposing both deterministic and random strategies for file placement under coded multicast.
\cite{Multihop} further studies the effectiveness of cooperative local caching under different file request distributions, network sizes and cache sizes with multi-hop communications.

Finally, there are also some works \cite{BoHanPanHui,ChenProulx,SocialCellular,SocialVTC} in the literature discussing the file or information sharing in wireless social networks \cite{SocialD2D}. They exploit the fact that people who are socially connected are more likely to share similar interests or have similar mobility patterns, which can be utilized for the file and information sharing in the wireless network. \cite{BoHanPanHui} discusses the  offloading of mobile data traffic through opportunistic communications between neighbouring MTs. With the social information between the mobile phones, a minimum target set is selected for the maximization of data offloading. \cite{ChenProulx} exploits the social information to devise a social-network-assisted cooperation for the device-to-device communications. {In \cite{SocialCellular}, a social group multicast scheme in supplement to the cellular network is proposed for delay-sensitive file transmissions. In contrast to \cite{SocialCellular} under  the multicast of BS, in \cite{SocialVTC}, a social multicast scheme between the MTs is proposed for wireless ad-hoc network exploiting the social network information to reduce the file transmission delay.}

Our work and those on wireless social network are similar in terms of file or information sharing between devices in the network.  However, the difference is that, in  the wireless social network,  the social information between the MTs  is required such that information dissemination or file sharing can achieve a better performance.  While in our work, we utilize the personal file preference information of the MTs for cooperative local caching instead of the social information between the MTs. { Moreover, although heterogeneous file preferences have been considered in the literature, such as \cite{JiConf} and \cite{ChenLZT16}, the heterogeneity of preferences in the cooperative environment still needs to be further investigated.} In this work, we investigate the performance trade-offs between different groups of MTs in heterogeneous file preference and the impacts of selfish behaviours in cooperation.

\section{System Model}\label{sec:system_model}

\begin{figure}[t]
  \centering
  \includegraphics[width=8cm]{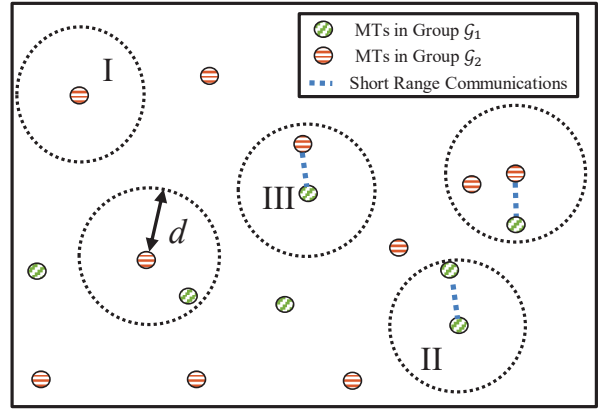}
  \caption{Cooperative local caching enabled file sharing among MTs in two groups. Different cases of file sharing: I. no file sharing, II. intra-group file sharing, and  III. inter-group file sharing.}\label{sysmodel}\vspace{-2em}
\end{figure}

\begin{table}[t]\caption{List of Symbols and Their  Meanings.}
\centering

\begin{tabular}{l  L{7cm}@{} }
\hline
Symbols& Meanings\\
\hline
$\mathcal{K}$, $K$ & Set of interest groups and number of groups\\
$\mathcal{G}_k$  &  Set of MTs in group $k\inm{K}$\\
$\lambda_k$, $\lambda_0$ & Spatial density of MTs in group $k$ and all the groups\\
$\Phi_k$& Poisson Point Process for the MTs' locations in group $k$\\
$X_{k,i}$& Coordinate of MT $i$ in group $k$\\
$\mu_{k}$, $\mu_0$  & Average number of MTs within distance $d$ in group $\mathcal{G}_k$ and all the groups\\
$d$& Range of short range communications\\
$\cF$ & Set of the files\\
$c_{k,f}$, $\reqprob$ & Group caching and request distribution for file  $f\inm{F}$ in group $k\inm{K}$\\
$C_{f}$,$R_f$& Social file caching and request distribution for file $f\inm{F}$\\
$\cU_k$, $\cU$ & Utility of group $\cG_k$ and the social utility\\
$\cR$& Feasible utility region\\
\hline
\end{tabular}
\label{tab:TableOfNotationForMyResearch}
\end{table}

{Before introducing the system model, we first summarize all the notations used in this paper in Table \ref{tab:TableOfNotationForMyResearch} for the ease of reading.} In this section, we introduce the system model of cooperative local caching. As illustrated in Fig. \ref{sysmodel}, we consider a large number of MTs served by the wireless cellular network and these MTs can also leverage short-range communications (e.g., Bluetooth \cite{bluetooth}) to share cached files (e.g., videos) with each other upon file requests during the peak hours. According to these MTs' file preferences, we practically categorize them into $K$ groups, denoted by the set $\cK=\{1,2,\cdots,K\}$. Fig. \ref{sysmodel} shows an example of two groups (i.e., $K=2$) in the network and also illustrates different cases of file sharing between the MTs. We denote the MTs in each group $k\inm{K}$ by $\cG_k$ and we assume that their locations follow a two-dimensional independent HPPP $\Phi_k=\{X_{k,i}\},~i\inm{G}_k$ \cite{Kingman1993} with spatial density $\lambda_k,~k\inm{K}$ and the MTs' coordinates $X_{k,i}\in\mR^2$, which are independent of the other groups. By the superposition theorem of Poisson process \cite{Kingman1993}, we also denote the {\it social density} of the MTs as the sum  of all the groups' spatial densities, i.e., $\lambda_0=\sum_{k\inm{K}}\lambda_k$.  For a certain MT in each group, it can potentially acquire its requested files from its {\it neighbouring MTs}  via short range  communications (in the same or different groups) within the distance $d$, as illustrated by the circle in Fig.~\ref{sysmodel}. Here, $d$ can be the range of communications in the wireless network. Specifically, let $B(T,x)\in\mR^2$ denote a disk of radius $x$ centered at $T$.   Then, considering a typical MT $i\inm{G}_k$ in group $k$,  $B(X_{k,i},d)$ denotes the area where file sharing is possible between the MT at location $X_{k,i}\in\Phi_k$ and all the other MTs both inside and outside the group $k$. Therefore, under the HPPP model, the average number of neighbouring MTs from group $k$ for MT $i$ in the area of $B(X_{k,i},d)$ is $\mu_k=\pi d^2\lambda_k,~k\inm{K}$ and the average number of neighbouring MTs from all the groups in  $B(X_{k,i},d)$ is $\mu_0=\sum_{k\inm{K}}\mu_k=\pi d^2\lambda_0$.

Cooperative local caching allows an MT to share the requested file locally with another MT (if any). However, this local file sharing highly depends on the caching strategies of all the $K$ groups of MTs.  We introduce the system model of the cooperative local caching in the following.}

\subsection{System Model of Cooperative Local Caching}\label{subsec:DeploymentPhase}
 During the off-peak hours, MTs cache files into their local memories. There are mainly two approaches for the file caching, namely {\it deterministic caching} (e.g., \cite{TITFemto}) and {\it random caching} (e.g., \cite{JiCM13}). Deterministic caching is generally the approach used in Femto-caching \cite{TITFemto}, where the locations of the Femto-BSs are fixed. However, this approach may not be suitable for cooperative local caching in our case due to the mobility of the MTs. In contrast to deterministic caching, random caching can  easily address the mobility issues \cite{JiCM13}. Therefore, in this paper, we consider random caching as our caching scheme. We define the set of popular files as $\cF=\{1,2,\cdots,F\}$ and assume that each MT caches one file from the library $\cF$.\footnote{We assume each MT only caches a single file for the tractability of analysis. However, our results can be extended to multi-file caching, which offers essentially the same insights.} It caches  file $f\inm{F}$ based on independent random sampling from the {\it group caching distribution} $\mv{c}_{k}=[c_{k,1},c_{k,2},\cdots,c_{k,F}]$, which is defined as the probability mass function (PMF) over the files $\cF$ with
 \begin{align}
   \sum_{f\inm{F}}c_{k,f}&=1,f\inm{F},\label{MemoryConstraint}\\
   0\leq c_{k,f}&\leq 1,~k\inm{K},~f\inm{F}.\label{NonNegative}
 \end{align}
 We also integrate the group caching distributions of the other groups $\cK\setminus\{k\}$ as $\mv{c}_{- k}=[\mv{c}_1,\cdots,\mv{c}_{k-1},\mv{c}_{k+1},\cdots,\mv{c}_K]$.   Then, given all the group caching distributions ($c_{k,f}, ~\forall k\inm {K},~f\inm{F}$) and the densities of the groups ($\lambda_k,~\forall k\inm {K}$), we denote the {\it social caching distribution} as the average group caching distributions with the weights equal to their normalized densities as
\begin{align}\label{SocialCaching}
C_f=\frac{1}{\lambda_0}\sum_{k\inm{K}}\lambda_k{c_{k,f}},~f\inm{F}.
\end{align}
$C_f$ has its physical meaning in our cooperative local caching scheme as it denotes the average availability of a certain file $f\inm{F}$ for a particular MT within unit area.

During the peak hours, MTs make requests to the files based on their preferences over the files. As we divide the MTs into different groups according to their interests, the MTs within the same group $\mathcal{G}_k$ have a common {\it group request distribution} $\mv{r}_k=[r_{k,1},r_{k,2},\cdots,r_{k,F}]$ with $\sum_{f\inm{F}}\reqprob=1$. We assume all the files have positive request probabilities (i.e., $r_{k,f}>0,~\forall k\inm{K}$). In a certain realization,  each MT in the group $\cG_k,~k\inm{K}$ makes an independent request to a certain file $f\inm{F}$ based on a random sampling from  $\mv{r}_k$.  We also denote $\mv{r}_{-k}=[\mv{r}_1,\cdots,\mv{r}_{k-1},\mv{r}_{k+1},\cdots,\mv{r}_K]$ as the aggregate group request distributions of the other groups $\cK\setminus\{k\}$. Finally, for the MTs in all the groups, we denote  the {\it social  request distribution} as the average group request distributions with the weights equal to their normalized densities, that is
\begin{align}\label{SocialReq}
R_f=\frac{1}{\lambda_0}\sum_{k\inm{K}}\lambda_k{r_{k,f}},~f\inm{F}.
\end{align}

Under the request of an MT in the group $\cG_k,~k\inm{K}$ for a particular file $f\inm{F}$, we are ready to specify the file download protocol as follows.
\begin{itemize}
    \item [1)]\emph{Obtaining the file from on-board cache:} If the file $f\inm{F}$ is already cached in its memory, the MT can obtain the file directly. {If the file is not found, the MT requests the file from the neighbouring MTs;}
    \item [2)] \emph{Obtaining the file from neighbouring MTs:}  The MT requests  the file from its neighbouring MTs within the range $d$. The neighbouring MTs can be from its own group or the other groups. If multiple neighbouring MTs have the file, the MT chooses one randomly.
\end{itemize}
Based on the above-specified system model of the cooperative local caching, in the next sub-section, we introduce the definition of utility for measuring the performance of each group.

\subsection{Definition of Utilities for  Cooperative Local Caching}
{With the above-specified transmission protocol, we are ready to define the utility function for characterizing the quality of service (QoS) for all the groups. For quantifying their performances, there may be various methods for the definitions of the utility functions. In cooperative local caching, we observe that each MT should try to increase the probability of {\it successful file discovery} on its own cache or its neighbouring MTs, which can benefit the MT or the network in various aspects. Some high-level examples of such benefits are given as follows:
\begin{itemize}
    \item {\it Delay reduction of the file download:} The delay of downloading from an MT in the range of local file sharing is usually much lower than that from the remote servers \cite{TITFemto}. Hence,  higher probability of file discovery results in a lower delay in the file download, which is beneficial for the QoS of the MTs.

    \item {\it Bandwidth usage and backhaul capacity reduction:} Due to short range of the file sharing, cooperative local caching can effectively reduce the bandwidth usage \cite{LocalCachingCommMag}. Moreover,  cooperatively sharing the file locally reduces the need for high-capacity backhaul in the wireless network\cite{TaoMeixiaYuWei}. Hence, higher file discovery probability reduces the bandwidth and backhaul usage, which is beneficial for the QoS of the whole network.
\end{itemize}
Based on the above discussions, we give two definitions of the utility function in the following.
\subsubsection{Group Utility}
First, we define the utility of each group as the probability of successful file discovery for the MTs in the group.  We define the event that, when requesting a certain file, the file is found in its own cache or that of more than one neighbouring MT as $E$. Then, the probability of successful file discovery for the MTs in group $k$ under the caching distributions $\mv{c}_k$ and those of all the other  groups $\mv{c}_{-k}$ is defined  as $\mP(E;\mv{c}_k,\mv{c}_{-k})$. Now, we are ready to specify the {\it group utility} of the MTs in group $k$   as the probability of successful {file discovery} in its own cache as well as any of the other MTs in the range of $d$ as
\begin{align}\label{utility_group}
\cU_k(\mv{c}_k;\mv{c}_{-k})=\mP(E;\mv{c}_k,\mv{c}_{-k}).
\end{align}
With the following theorem, we provide the closed-form expression for the group utility function.
\begin{theorem}\label{SysModel:delay}
Given the group caching distribution $\mv{c}_k$ and those of the other groups $\mv{c}_{-k}$, the utility for the MTs in  group $k\inm{K}$ is
\begin{align}
&\cU_k(\mv{c}_k;\mv{c}_{- k})=\sum_{f\inm{F}}\reqprob\left(1-\bar{c}_{k,f}e^{-\mu_0C_{f}}\right),\label{result2}
\end{align}
where $e^{-\mu_0C_{f}}$ with $C_f$ given in (\ref{SocialCaching}),  denotes the probability that file $f\inm{F}$ is not  found in any neighbouring MT within the range $d$, and ${\bar{c}_{k,f}}=1-{\cacheprob}$ denotes the complement of the group caching distribution.

\end{theorem}
 \begin{proof}
 Please refer to Appendix \ref{Appendix:proofthm1}.
 \end{proof}

\begin{remark}
From (\ref{result2}) in Theorem \ref{SysModel:delay}, we can observe that, for a certain group $k$, its utility relies on its own group request distribution ${r}_{k,f}$ and caching distributions  ${c}_{k,f}$, as well as the social density $\lambda_0$ ($\mu_0=\pi d^2\lambda_0$) and the social caching distribution ${C}_f$, but regardless of the specific caching distribution of the other groups $\mv{c}_{-k}$ or their spatial densities.  We can also observe that, for a certain file $f\inm{F}$, the group utility $\cU_k(\mv{c}_k;\mv{c}_{- k})$ for the MTs in group ${\cG}_k$ monotonically increases with respect to the group caching distribution $\cacheprob$ and the social caching distribution $C_f$. This is because if the MT or its neighbouring MTs cache the file with higher probability, the requested file is more available to the neighbouring MT and thus the utility increases. Moreover, from (\ref{result2}),  the effect of the file not cached in its own MT (i.e., $\bar{c}_{k,f}$) is exponentially vanishing with respect to the social density $\lambda_0$ and caching distribution $C_f$. This shows how the neighbouring MTs' caches and the MT's own cache compliment with each other under our mobility model.
\end{remark}

\subsubsection{Social Utility}
Apart from the above definition of group utility, another definition of the utility function is based on the performance of the whole network, which is defined as the weighted-sum of each group's utility defined in (\ref{utility_group}) with respect to the normalized density of each group in the network as
\begin{align}\label{SysModel:SocialUtility}
\cU(\{\mv{c}_k\})=\frac{1}{\lambda_0}\sum_{k\inm{K}}{\lambda_k}\cU_k(\mv{c}_k;\mv{c}_{-k}).
\end{align}
The utility function reflects the relative importance of the groups in the whole society. Hence, this definition of utility can be termed as the {\it social utility} of the network.}

Next, in order to explicitly show the trade-offs between different groups under cooperative local caching, we define the feasible utility region of different groups.
\subsection{Feasible Utility Region}
For our proposed cooperative local caching scheme, we intend to increase the utilities of all the groups under the constraint for the allocations of caching distribution over the files. However, different groups have conflicts of interests under heterogeneous file preferences. For example, consider a certain file that is not popular in one group but popular in the other groups. If the former group caches this file with high probability, the utilities of the other groups will increase because of the file's popularity. However, this will reduce the probabilities for the former group to  cache popular files and thus decrease its utility. In order to explicitly characterize such conflicts of interests, we define the {\it feasible utility region} of all the groups as follows.
\begin{definition}
The {\it feasible utility region} of different groups is the set of utility tuples that different groups can achieve simultaneously, which is given as
\begin{align}\label{delayregion}
\cR=\bigcup_{(\ref{MemoryConstraint}),(\ref{NonNegative})}\begin{cases}(u_1,u_2&\hspace{-1em},\cdots,u_K):~u_k\leq \cU_k(\mv{c}_k;\mv{c}_{-k}),~k\inm{K}
\end{cases}\Big\}.
\end{align}
\end{definition}
\begin{figure}[t]
  \centering
  \includegraphics[width=8cm]{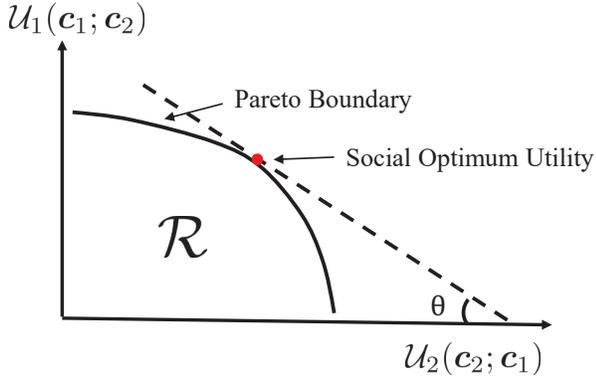}
  \caption{Feasible utility region, Pareto boundary and social optimum with slope $\theta=\arctan{\frac{\lambda_1}{\lambda_2}}$ for two groups.}\label{ParetoRegion}
\end{figure}
This region denotes the {\it Pareto region} of all the possible utilities under different allocations of group caching distributions $\mv{c}_k,~k\inm{K}$. An example of the feasible utility region is given in Fig. \ref{ParetoRegion}, which shows the trade-off between the utilities of two groups ($K=2$). Within the feasible utility region, the {\it Pareto optimality} and {\it Pareto boundary} are defined as follows.
\begin{definition}
A utility tuple $(u_1,u_2,\cdots,u_K)\inm{R}$ is {\it Pareto optimal} if there is no other utility tuple $(u_1',u_2',\cdots,u_K')\inm{R}$ with $(u_1',u_2',\cdots,u_K')\geq (u_1,u_2,\cdots,u_K)$, where the inequality is component-wise. The set of tuples in the utility region that are Pareto optimal are called the {\it Pareto boundary} of the feasible utility region $\cR$.
\end{definition}

In order to characterize the complete trade-offs between different groups of MTs, in the following section, we consider the scenario of full cooperation under centralized coordination in our cooperative local caching scheme.
\section{Fully Cooperative local Caching}\label{sec:FullCoop}
In this section, we consider the case where different groups $\cG_k,~k\inm{K}$ are {\it fully cooperative} and follow a centralizer's instructions to cache the files for the benefits of the whole society. It should be noted that the preferences of the MTs in each group can be readily available on the service provider's platform (such as YouTube, Netflix, etc.). The optimization of the group caching distribution can also be easily performed in the centralizer where the computation resources are abundant.  In order to completely characterize the feasible  utility region $\cR$ defined in (\ref{delayregion}), we first define the weighted-sum utility of different groups as
\begin{align}
\cU(\{\mv{c}_k\},\mv{w})=\sum_{k\inm{K}}w_k\cU_k(\mv{c}_k;\mv{c}_{-k}),
\end{align}
where $\cU_k(\mv{c}_k;\mv{c}_{-k})$ is the utility for group $k$ defined in (\ref{result2}) and $\mv{w}=[w_1,w_2,\cdots,w_K]$, with the weights $w_k\geq 0,~k\inm{K}$ subjected to $\sum_{k\inm{K}}w_k=1$, represents the relative importance of different groups in the society.  More specifically, given group $k$'s utility, the {social utility} defined in (\ref{SysModel:SocialUtility}) can be obtained by equating weight $w_k$ to group $k$'s normalized spatial density (i.e., $w_k=\lambda_k/\lambda_0$). We then propose the following weighted-sum utility maximization problem for deciding the $\mv{c}_k$'s for all the groups.
\begin{subequations}
\begin{align}
\mathrm{(P1)}:~\mathop{\mathtt{max.}}_{\{\mv{c}_k\}}&~~ \cU(\{\mv{c}_k\},\mv{w}) \nonumber\\
\mathtt{s.t.}
&~~\sum_{f\inm{F}}{\cacheprob}= 1,~k\inm{K},\label{p0robust}\\
&~~0\leq {\cacheprob}\leq 1,~f\inm{F},~k\inm{K}.\label{p0integer}
\end{align}
\end{subequations}
Among all the Pareto optimal utility tuples, the {\it social optimal utility} is defined as the optimal solution to problem $\mathrm{(P1)}$ with $w_k=\lambda_k/\lambda_0,~k\inm{K}$ in the objective function $\cU(\{\mv{c}_k\},\mv{w})$,  corresponding to the social utility in (\ref{SysModel:SocialUtility}). An example is shown in Fig. \ref{ParetoRegion} for the case of $K=2$, where the social optimum is the intersection between the line with slope $\arctan\frac{\lambda_1}{\lambda_2}$ and the Pareto boundary.

However, it can be verified that the objective function in problem $\mathrm{(P1)}$ is non-concave  due to the coupling terms in $C_f=\frac{1}{\lambda_0}\sum_{k\inm{K}}\lambda_kc_{k,f}$ and thus the global optimum solution is hard to obtain. Although the problem is not jointly concave with respect to all the $c_{k,f},~k\inm{K},~f\inm{F}$, it can be proved that the objective function $\cU(\{\mv{c}_k\},\mv{w})$ is marginally concave with respect to the caching distribution $\mv{c}_k$ for a certain group $k\inm{K}$. This is shown in the following lemma.
\begin{lemma}\label{MarginalConvexLemma}
The weighted-sum utility $\cU(\{\mv{c}_{k}\},\mv{w})$ in problem $\mathrm{(P1)}$ is marginally concave with respect to the group caching distribution $\mv{c}_{k}$ with the group caching distributions of the other groups $\mv{c}_{-k}$ fixed.
\end{lemma}
\begin{proof}
We obtain the Hessian matrix of objective function of problem $\mathrm{(P1)}$ with respect to $\mv{c}_k$
\begin{align}
\frac{\partial^2\cU(\{\mv{c}_k\},\mv{w})}{\partial\bar{c}_{k,f}^2}&=
-\Bigg(w_kr_{k,f}e^{-\mu_0C_f}\mu_k(2+\bar{c}_{k,f}\mu_k)\nonumber\\
&+\sum_{j\inm{K}\setminus\{k\}}w_jr_{j,f}\bar{c}_{j,f}e^{-\mu_0C_f}\mu_k^2\Bigg),
\\
\frac{\partial^2\cU(\{\mv{c}_k\},\mv{w})}{\partial\bar{c}_{k,f}\partial\bar{c}_{k,f'}}&=0,
\end{align}
where $f'\inm{F}\setminus\{f\}$. By the negative diagonal dominance, its negative semi-definiteness is proved.
\end{proof}

The above result not only shows its concavity, but also reveals the decomposition structure of the problem with respect to ${c}_{k,f},~k\inm{K},~f\inm{F}$. Based on this, we propose a coordinate descent algorithm \cite{boyd2004convex} for  problem $\mathrm{(P1)}$, which sequentially optimizes the weighted-sum utility function with respect to different groups. For the coordinate descent algorithm, special attentions need to be given to those groups with zero weights. We divide the whole set of groups $\cK$ into two sets $\cK_+$ and $\cK_0$ with $\cK_+\cup \cK_0=\cK$ and $\cK_+\cap\cK_0=\emptyset$.  $\cK_+\subseteq\cK$ denotes those groups with positive weights such that $w_k>0,~\forall k\inm{K}_+$ and $\sum_{k\inm\cK_+}w_k=1$; while $\cK_0\subset\cK$ denotes those groups with zero weights such that $w_k=0,\forall~k\inm{K}_0$.

First, we consider the groups with positive weights $k\inm{K}_+$. Given the caching distributions of the other groups $\mv{c}_{-k}$, for a certain group $k\inm{K}_+$, we aim to solve for the following problem.
\begin{subequations}
\begin{align}
\mathrm{(P2-1)}:~\mathop{\mathtt{max.}}_{\mv{c}_k}&~~ \sum_{j\inm{K}_+}w_j\sum_{f\inm{F}}r_{j,f}\left(1-\bar{c}_{j,f}e^{-\mu_0C_{f}}\right) \nonumber\\
\mathtt{s.t.}
&~~\sum_{f\inm{F}}{\cacheprob}= 1,\label{p2robust}\\
&~~0\leq {\cacheprob}\leq 1,~f\inm{F}.\label{p2integer}
\end{align}
\end{subequations}
Thanks to the marginal concavity shown in Lemma \ref{MarginalConvexLemma}, we derive the optimal solution for the above problem as given in the following theorem by leveraging its KKT conditions \cite{boyd2004convex}.
\begin{theorem}\label{SocialTheorem}
The optimal group caching distribution of group $k$ in problem $\mathrm{(P2-1)}$ is
\begin{align}\label{thmSocial:result}
{c_{k,f}^{*}}=\left[1-\Bigg(\frac{1}{\mu_k}\cW\Big(A_{k,f}e^{\mu_kB_{k,f}}\Big)-{B_{k,f}}\Bigg)\right]_0^1,
\end{align}
 with
 \begin{align}
 A_{k,f}&=\frac{\gamma_k^*}{w_ke^{-\mu_0}r_{k,f}e^{\sum_{j\inm{K}\setminus\{k\}}\mu_j\bar{c}_{j,f}}},\label{Thma}\\
 B_{k,f}&=\frac{1}{\mu_k}+\sum_{j\inm{K}_+\setminus\{k\}}\frac{w_jr_{j,f}}{w_kr_{k,f}}\bar{c}_{j,f},\label{Thmb}
 \end{align}
where $[\cdot]_0^{1}=\min\{1,\max\{0,\cdot\}\}$, $\cW(\cdot)$ is the Lambert-W function \cite{corless1996lambertw} and $\gamma^*_k$, which is the optimal dual variable of constraint (\ref{p2robust}), denotes a constant that satisfies $\sum_{f\inm{F}}{{c}_{k,f}^*}=1$.
\end{theorem}
\begin{proof}
Please refer to Appendix \ref{sec:proofthm2}.
\end{proof}

Next, for the groups ${k\in\cK_0}$ with zero weights, they cache completely un-selfishly for the groups with positive weights in ${\cK_+}$. In this case, the weighted-sum utility for the optimization of the group $k\inm{K}_0$ given the caching distributions $\mv{c}_{-k}$ of the other groups reduces to
\begin{align}
\sum_{k\inm{K}}w_k\cU_k(\mv{c}_k;\mv{c}_{-k})=1-\sum_{f\inm{F}}e^{\mu_k\bar{c}_{k,f}+D_{k,f}},
\end{align}
where $D_{k,f}$ is given by
\begin{align}
&D_{k,f}=\ln\left[\left(\sum_{j\inm{K}_+}w_jr_{j,f}\bar{c}_{j,f}\right)e^{-\mu_0}e^{\sum_{l\inm{K}\setminus\{k\}}\mu_l\bar{c}_{l,f}}\right].
\end{align}
Then, we  need to solve the following problem.
\begin{subequations}
\begin{align}
\mathrm{(P2-2)}:~\mathop{\mathtt{max.}}_{\mv{c}_k}&~~ 1-\sum_{f\inm{F}}e^{\mu_k\bar{c}_{k,f}+D_{k,f}} \nonumber\\
\mathtt{s.t.}
&~~\sum_{f\inm{F}}{c_{k,f}}= 1,\label{p3robust}\\
&~~0\leq c_{k,f}\leq 1,~f\inm{F}.\label{p3integer}
\end{align}
\end{subequations}
Because the function $e^x$ is convex and linear combination preserves convexity \cite{boyd2004convex}, the objective in problem $\mathrm{(P2-2)}$ is a concave function. The constraints are also affine. Therefore,  problem $\mathrm{(P2-2)}$ is a convex optimization problem and its global optimum can be obtained in the following theorem.
\begin{theorem}\label{zeroTheorem}
The optimal solution to problem $\mathrm{(P2-2)}$ is given by
\begin{align}
c_{k,f}^*=\left[1-\frac{1}{\mu_k}\left(\ln\left(\frac{\gamma_k^*}{\mu_k}\right)-D_{k,f}\right)\right]_0^1,\label{zeroresult}
\end{align}
where $\gamma^*_k$ is the optimal dual variable for constraint (\ref{p3robust}), which can be obtained by substituting the result in (\ref{zeroresult}) into the equation $\sum_{f\inm{F}}c_{k,f}^*=1$.
\end{theorem}
\begin{proof}
The proof is similar to that of Theorem \ref{SocialTheorem} and we provide the sketch of both proofs as follows: we first need to derive the dual problem and prove the zero duality gap between the primal and dual problems. We then need to decompose the original problem with respect to each file via primal decomposition. Finally, we need to derive the optimal caching distribution with the KKT conditions.
\end{proof}
Now, we have obtained the optimal solutions to problems $(\mathrm{P2-1})$ and $(\mathrm{P2-2})$ for the groups with positive and zero weights, respectively. Based on the above results,  we propose Algorithm I in Table \ref{algorithm:1} for problem $(\mathrm{P1})$ based on the coordinate descent method \cite{boyd2004convex}. Because the feasible set of the problem is a compact set and the objective function is lower-bounded, the algorithm is guaranteed to converge to at least a local maximum of problem $\mathrm{(P1)}$.

\begin{table}[t]
\begin{center}
\caption{Coordinate Descent Algorithm for Solving Problem $\mathrm{(P1)}$ under Full Cooperation.}
 \hrule\vspace{0.2cm} \textbf{Algorithm I}   \vspace{0.2cm}
\hrule
\begin{itemize}
\item [1.]{\bf Repeat for} $k=1,2,\cdots,K$ in iterations $i=1,2,\cdots$
\begin{itemize}
\item[1)]
{\bf Initialize:} $\gamma^{(l)}_k \coloneqq 0$, $\gamma^{(h)}_k\coloneqq \infty$;
\item[2)]
{\bf Repeat:}\begin{itemize}
    \item [i.]$\gamma_k\coloneqq\frac{1}{2}(\gamma_k^{(l)}+\gamma_k^{(h)})$;
    \item [ii.] {\bf If} $k\inm{K}_+$, update $c_{k,f}^*$ according to (\ref{thmSocial:result}) in Theorem \ref{SocialTheorem};\\
     {\bf If} $k\inm{K}_0$, update $c_{k,f}^*$ according to (\ref{zeroresult}) in Theorem \ref{zeroTheorem};
    \item[iii.] {\bf If} $\sum_{f\inm{F}}c_{k,f}^*< 1$, set $\gamma_k^{(h)}\coloneqq \gamma_k$; {\bf Else}, set $\gamma_k^{(l)}\coloneqq \gamma_k$;
        \end{itemize}
\item[3)] {\bf Until:} the condition $|\gamma_k^{(l)}-\gamma_k^{(h)}|>\delta_\gamma$ is violated.
\end{itemize}
\item [2.]  {\bf Until:} $|\cU^{i}(\{\mv{c}^*_k\})-\cU^{i-1}(\{\mv{c}^*_k\})|\leq \delta_{\cU}$.
\end{itemize}
\vspace{0.1cm} \hrule \label{algorithm:1}
\end{center}
\end{table}

\section{Partially Cooperative and Non-cooperative local Caching}\label{sec:PartialCoopNoCoop}
In the previous section, we have proposed a centralized algorithm under full cooperation to achieve the approximate Pareto boundary of the feasible utility region. { Such approximation is due to the difficulty in proving the convexity of the utility region and optimally solving the non-convex problem $\mathrm{(P1)}$}. While different groups may be fully cooperative and their caching decisions can be jointly optimized, this scheme does not apply to the case where different groups have selfish behaviours.  In this section, we discuss two possible benchmark schemes by investigating some groups' selfish behaviours under {\it partial cooperation} or {\it no cooperation}. We first consider the  partially cooperative case that still allows {\it inter-group file sharing} (illustrated in Fig. \ref{sysmodel}), while different groups are not willing to achieve a common social utility as in the case of full cooperation and the caching decisions are made independently by each group. Next, we consider the non-cooperative caching, where there is only {\it intra-group file sharing} (illustrated in Fig. \ref{sysmodel}) but no longer inter-group file sharing.

\subsection{Partially Cooperative Caching with Inter-group File Sharing}\label{subsec:PartialCoop}
We assume that each group only knows its own group's file request distribution $\mv{r}_k$ and the social file request distribution $R_f$, which is public information that can be accessed by all the groups when making caching decisions.\footnote{This is possible when the MTs are under the same platform (such as YouTube, Netflix, etc.), where the statistics of the preference of the whole society can be publicly available. However, due to the issues of privacy, the exact preference of a certain group may not be available to the other groups.} Based on the above information, each group makes its own decision in caching. Different groups of MTs are partially cooperative such that they are able to share files with each other, but they are also selfish in their own caching strategy due to the heterogeneous file preferences and aim to increase their own utility.  In this case, not knowing the exact caching distribution of the other groups, it is reasonable for each group to assume that the other groups are faithful to their own preferences and they cache files according to their  preferences, (i.e.,  $\mv{c}_{-k}=\mv{r}_{-k}$). Then, the group utility $\cU_k(\mv{c}_k;\mv{c}_{-k})$ in (\ref{result2}) reduces to
\begin{align}
&\cU_k(\mv{c}_k;\mv{r}_{-k})\nonumber\\
&=\sum_{f\inm{F}}\reqprob \bigg(1-\bar{c}_{k,f}e^{-\mu_0R_{f}+\mu_kr_{k,f}-\mu_kc_{k,f}}\bigg).
\end{align}
{The above expression is derived by substituting $\mv{c}_{-k}=\mv{r}_{-k}$ into the group utility function defined in (\ref{utility_group}) and utilizing the formula $R_f=1/\lambda_0\sum_{k\inm{K}}\lambda_kr_{k,f}$.}
From the above result, we can see that, under the assumption of $\mv{c}_{-k}=\mv{r}_{-k}$ for a certain group $k$, it does not need to know the actual file request distributions of the other groups $\mv{r}_j,~j\neq k$ to obtain its own utility $\cU_k(\mv{c}_k;\mv{r}_{-k})$, but only needs to know the aggregate social request distribution $R_f$ defined in (\ref{SocialReq}).

Given the above derivation of the utility of group $k$, in the following we aim to answer the following question: how should a group optimally cache files to maximize its own utility by exploiting the social file preference information $R_f$?  We formulate the above decision making for group $k$ as the following optimization problem that maximizes its utility given the caching distributions of the other groups as $\mv{c}_{-k}=\mv{r}_{-k}$.
\begin{subequations}
\begin{align}
\mathrm{(P3)}:~\mathop{\mathtt{max.}}_{\mv{c}_k}&~~ \sum_{f\inm{F}}\reqprob \bigg(1-\bar{c}_{k,f}e^{-\mu_0R_{f}+\mu_kr_{k,f}-\mu_kc_{k,f}}\bigg) \nonumber\\
\mathtt{s.t.}
&~~\sum_{f\inm{F}}{\cacheprob}= 1,\label{robust}\\
&~~0\leq {\cacheprob}\leq 1,~f\inm{F}.\label{integer}
\end{align}
\end{subequations}
The objective function is concave and this can be verified by checking Lemma \ref{MarginalConvexLemma} with $w_j=0,~\forall j\inm{K}\setminus\{k\}$ and $\mv{c}_{-k}=\mv{r}_{-k}$.  With all the constraints of this problem being affine,  problem $\mathrm{(P3)}$ is a convex optimization problem. { The feasible region constrained by (\ref{robust}) and (\ref{integer}) is a probability simplex and cannot be empty. Therefore, problem $\mathrm{(P3)}$ is always a feasible problem.} In the following theorem, we provide the closed-form solution for problem $\mathrm{(P3)}$ by checking its KKT conditions.
\begin{theorem}\label{theorem2}
Group $k$'s optimal group caching distribution for file $f\inm{F}$ under the partial cooperation  is
\begin{align}\label{thm2:result}
{{c}_{k,f}^*}=\Bigg[1-\frac{1}{\mu_k}\Bigg\{\cW\left(\frac{\gamma^*_ke^{1+\mu_0}}{r_{k,f}e^{\mu_0\bar{R}_f-\mu_k\bar{r}_{k,f}}}\right)-1\Bigg\}\Bigg]_0^1.
\end{align}
where   $\bar{R}_{f}=1-R_{f}$,  $\bar{r}_{k,f}=1-r_{k,f}$ and $\gamma^*_k$, which is the optimal dual variable of constraint (\ref{robust}), denotes a constant that satisfies $\sum_{f\inm{F}}{{c}_{k,f}^*}=1$.
\end{theorem}

\begin{proof}
The proof is similar to that of Theorem \ref{SocialTheorem} and is thus omitted here.
\end{proof}

In Theorem \ref{theorem2}, the allocation of caching distribution can be interpreted as water-filling over different files with $\gamma^*_k$ being the optimal water-level satisfying (\ref{robust}). Also, please be noted that in the above partial cooperation scheme, group $k\inm{K}$ only needs to know about the social file request distribution $R_f,~f\inm{F}$ instead of the specific request distributions of all the other groups $\mv{r}_{j},~j\inm{K}\setminus\{k\}$. Hence, the amount of information needed in partial cooperation is manageable in the individual decision of each group. Based on the above optimal solution, we propose  Algorithm II based on the bi-section method \cite{boyd2004convex} for problem $\mathrm{(P3)}$ in the case of partial cooperation in Table \ref{algorithm:2}.

\begin{table}[t]
\begin{center}
\caption{Optimal Algorithm for Problem $\mathrm{(P3)}$ in Partial Cooperation.}
 \hrule\vspace{0.2cm} \textbf{Algorithm II}   \vspace{0.2cm}
\hrule
\begin{itemize}
\item[1.] {\bf Initialize:} $\gamma^{(l)}_k \coloneqq 0$, $\gamma^{(h)}_k\coloneqq \infty$;
\item[2.] {\bf Repeat:}
    \begin{itemize}
    \item [1)]$\gamma_k\coloneqq\frac{1}{2}(\gamma^{(l)}_k+\gamma^{(h)}_k)$;
    \item [2)] ${{c}_{k,f}}=\Bigg[1-\frac{1}{\mu_k}\Bigg\{\cW\left(\frac{\gamma_k}{r_{k,f}e^{\mu_0\bar{R}_f-\mu_k\bar{r}_{k,f}}}\right)-1\Bigg\}\Bigg]_0^1$,
    \item[3)] {\bf If} $\sum_{f\inm{F}}c_{k,f}< 1$, set $\gamma^{(h)}_k\coloneqq \gamma$; {\bf Else}, set $\gamma^{(l)}_k\coloneqq \gamma$;
        \end{itemize}
\item[3.] {\bf Until:} the condition $|\gamma^{(l)}_k-\gamma^{(h)}_k|>\delta_\gamma$ is violated.
\end{itemize}
\vspace{0.1cm} \hrule \label{algorithm:2}
\end{center}
\end{table}

In the following, we give illustrative numerical examples to show the impacts of different system parameters on the optimal group caching distribution $c_{k,f}^*$ in Theorem \ref{theorem2} for the case of partial cooperation.
\begin{figure*}[t]
\centering
\begin{subfigure}{.5\textwidth}
  \centering
  \includegraphics[width=9cm]{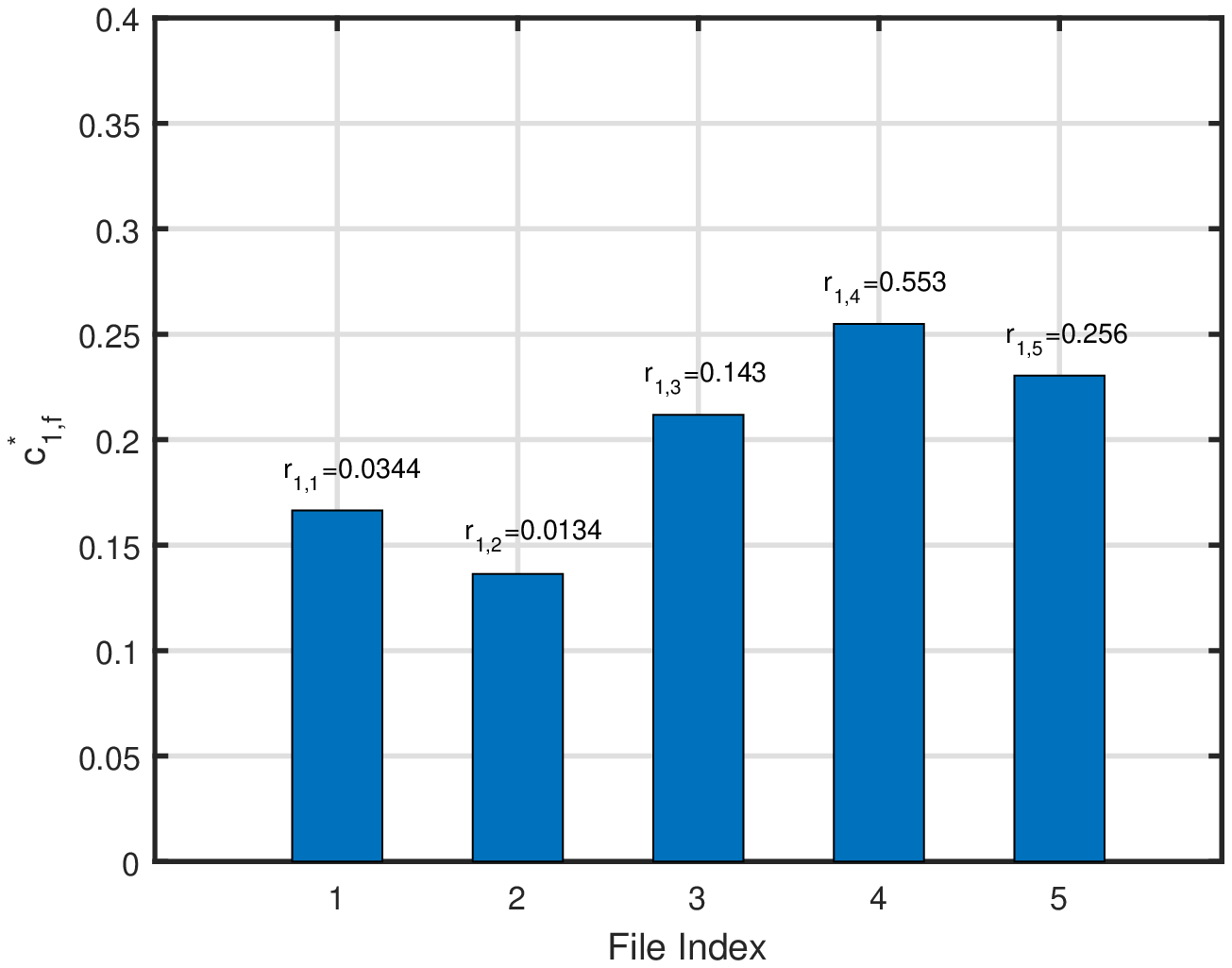}
  \caption{ }\label{comparefig1}
\end{subfigure}%
\begin{subfigure}{.5\textwidth}
  \centering
  \includegraphics[width=9cm]{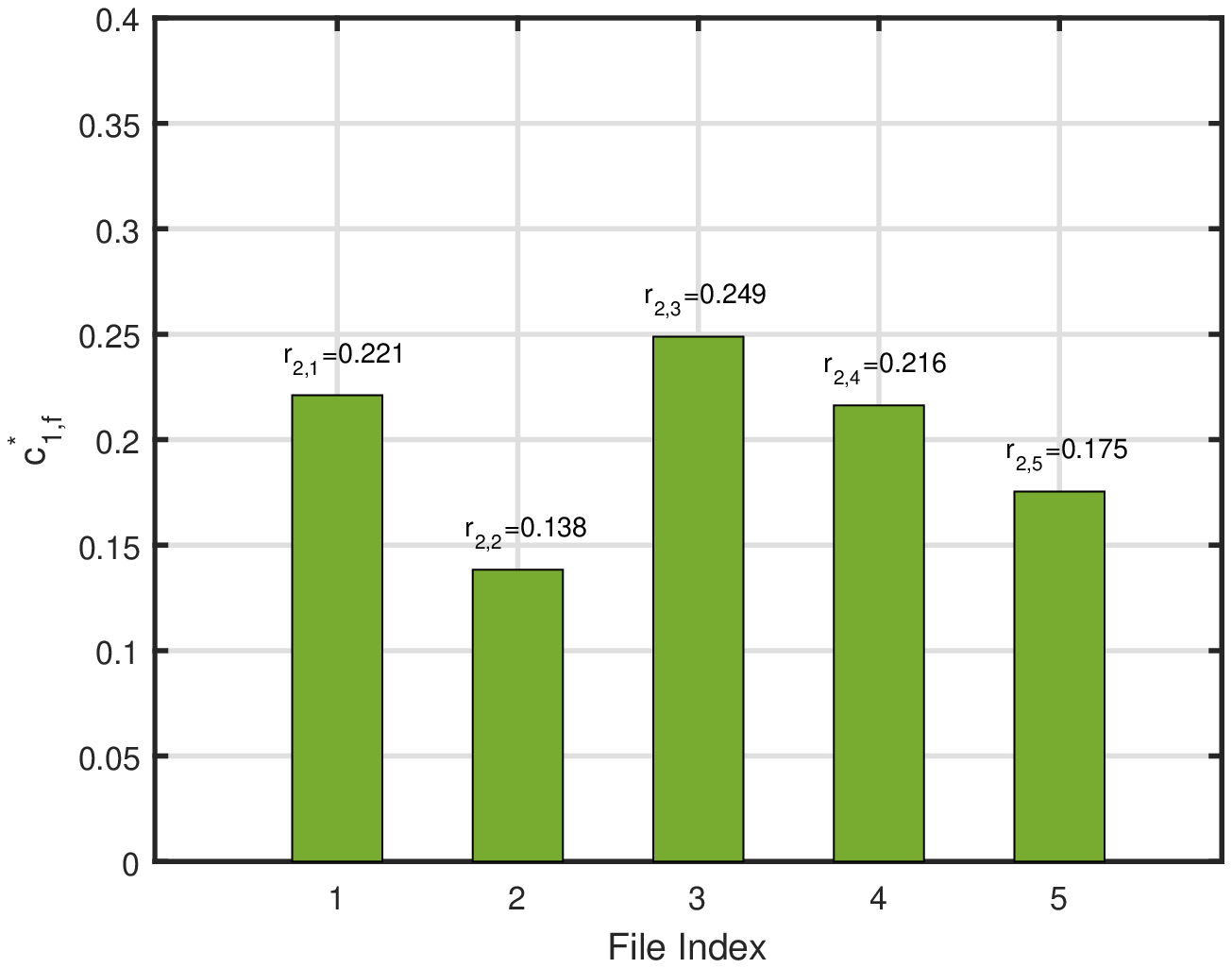}
  \caption{ }\label{comparefig2}
\end{subfigure}
\caption{Group 1's optimal caching distribution $c_{1,f}^*$  under (a) group 1's request distribution $r_{1,f}$, and (b) group 2's request distribution $r_{2,f}$.}\label{ProbAllocation}
\end{figure*}

\begin{example}\label{example1}
The settings for this example are given as follows: there are $K=2$ groups and $F=5$ files in total. The densities of MTs in groups 1 and 2 are $\lambda_1=0.05$ and $\lambda_2=0.05$ MTs per unit area, respectively. The range of communications is $d=5\mathrm{m}$\footnote{For example, the typical range of Class 2 Bluetooth is $5-10\mathrm{m}$\cite{bluetooth}.}. We separately examine the effects of one of the system parameters $r_{1,f}$ and $r_{2,f}$ on the optimal caching distribution $c_{1,f}^*$ of group 1 in Theorem \ref{theorem2}  while keeping the others equal across the 5 files.  First, for the equal values of these parameters across the 5 files: The group request distributions are uniform distributions with $r_{1,f}=1/5,~\forall f\inm{F}$ and $r_{2,f}=1/5,~\forall f\inm{F}$. Next, for the system parameters whose effects we want to examine: The PMFs  $\mv{r}_1$ and $\mv{r}_2$  are randomly generated with $\sum_{f\inm{F}}r_{1,f}=1$ and $\sum_{f\inm{F}}r_{2,f}=1$ for all the files $f\inm{F}$. The result is shown in Fig. \ref{ProbAllocation}. It can be observed from Fig. \ref{comparefig1} that the optimal caching distribution $c_{1,f}^*$ has water-filling structure \cite{goldsmith2005wireless} with respect to the request distributions of group 1 and group 2: $c_{1,f}^*$'s are monotonically increasing and decreasing in $r_{1,f}$ and $r_{2,f}$, respectively.  This is  because if the group request probability $r_{1,f}$ is high, it is desirable to match the caching distribution to the request distribution to increase the utility. On the other hand, it can also be observed from Fig. \ref{comparefig2} that,  if group 2 caches the file with high probability due to high $r_{2,f}$, there is less need for group 1 to cache the file and group 1 can exploit group 2 by caching the file with a lower probability.
\end{example}

\subsection{Non-cooperative Local Caching without Inter-group File Sharing}
Next, we consider the benchmark case of no cooperation, where different groups cannot share files with the other groups, but file sharing is still possible inside the group.  {This situation can be possible when different groups of MTs belong to different self-enclosed social communities, such as families, schools, etc., and can only share files within the group. With this benchmark case, we can examine the gain due to inter-group cooperation, either in the case of full cooperation or partial cooperation. The optimal caching distribution of a certain group $k\inm{K}$ can be seen as a special case of Theorem \ref{theorem2} with $\lambda_k>0$ and $\lambda_j=0,~j\inm{K}\setminus\{k\}$, which is equivalent to the result when there is only one group in the society in the case of full/partial cooperation.} The optimal caching distribution is specified in the following corollary.
\begin{corollary}\label{CorollaryNonCoop}
The optimal group caching distribution for group $k\inm{K}$ in the non-cooperative local caching scheme is
\begin{align}
{{c}_{k,f}^*}=\Bigg[1-\frac{1}{\mu_k}\Bigg\{\cW\left(\frac{\gamma^*_ke^{1+\mu_k}}{r_{k,f}}\right)-1\Bigg\}\Bigg]_0^1,
\end{align}
where $\gamma^*_k$ denotes a constant that satisfies $\sum_{f\inm{F}}{{c}_{k,f}^*}=1$.
\end{corollary}
Please be noted that the above result is also optimal for the special case of homogeneous file preference (i.e., $\mv{r}_k=\mv{r}_j,~\forall j,k\inm{K}$) where all the MTs belong to one group.

{
\section{Numerical Results}\label{sec:numerical}
In this section, we first validate the performance of the system with various parameters. We then proceed to the network simulation under the evolution of the system. The general simulation set-up is given for all the following simulations unless specified otherwise: for the group request distributions, we assume that they follow Zipf distribution as in \cite{ZipfDistribution}
\begin{align}\label{Zipf}
r_{k,f}=\frac{f^{-\gamma_r}}{\sum_{i=1}^Fi^{-\gamma_r}},~f\inm{F},~k\inm{K},
\end{align}
where  $\gamma_r$ is denoted as the {\it Zipf exponent}, measuring the concentration of the file popularity. The typical Zipf exponent with variations in different sources is around $\gamma_r=0.9$ according to the measurements in \cite{ZipfDistribution} and we set  $\gamma_r=0.9$ in all the simulations. { Similar to \cite{ChenLZT16}, in order to simulate the heterogeneous preferences of different groups, we randomly permute the request distribution of one group and assign them to the other groups.}  The range of communications is $d=5\mathrm{m}$ and the total number of files is $F=100$.

\subsection{Simulation under Two Groups}
In this sub-section, we provide numerical results for evaluating the performance of cooperative local caching under full, partial and no cooperation { in the simulation setup of two groups ($K=2$)}.

\subsubsection{Convergence of the Coordinate Descent Algorithm under Various Initializations}

{First, we show the convergence of Algorithm I under a simplified simulation setup with the number of files $F=5$.} For the approximate algorithm based on coordinate descent, although the convergence can be guaranteed, different initializations may lead to different local maximums due to the non-convexity of the problem. Hence, we investigate the impacts of different initialization methods, which are specified in the following.
\begin{itemize}
    \item [1.] {\bf Uniform Initialization} {\it (Unif. Init.)}: In uniform initialization, the initial caching distribution is uniform across the files regardless of their own or the other groups' request distributions: $\mv{c}_1=\frac{1}{F}\mv{1}^T_F,~\mv{c}_2=\frac{1}{F}\mv{1}^T_F$, where $\mv{1}_F$ denotes a vector of all ones with length $F$.
    \item [2.] {\bf Adaptive Initialization} {\it (Adapt. Init.)}: In adaptive initialization, the basic idea is to give the groups with higher weights more favourable initial distributions:
    $\mv{c}_1=\mv{r}_1,~\mv{c}_2=\mv{r}_1, \mathrm{if~} 0.66\leq w_1\leq 1$,
    $\mv{c}_1=\mv{r}_1,~\mv{c}_2=\mv{r}_2, \mathrm{if~} 0.33\leq w_1< 0.66$ and
    $\mv{c}_1=\mv{r}_2,~\mv{c}_2=\mv{r}_2, \mathrm{if~} 0 \leq w_1 < 0.33$.
    \item[3.] { {\bf Optimal Randomized Initialization} {\it (Opt. Rand. Init.)}: In randomized initialization, the initial distribution is randomly generated with $\sum_{f\inm{F}}c_{1,f}=1$ and $\sum_{f\inm{F}}c_{2,f}=1$ and Algorithm I under such initialization is performed for 20,000 times. The convergence for the initialization with the highest converged utility is shown.}
\end{itemize}
\begin{figure}[t]
  \centering
  \includegraphics[width=9cm]{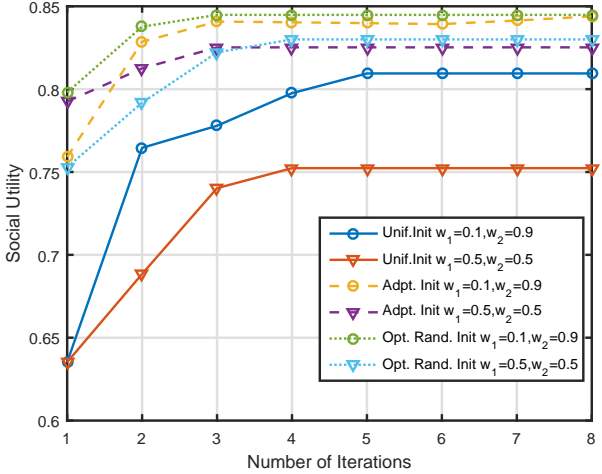}
  \caption{Convergence of the coordinate descent algorithm under various initialization methods.}\label{UnifCacheConvergence}
\end{figure}

The simulation results are shown in Fig. \ref{UnifCacheConvergence} with two different sets of weights $w_1=0.1$, $w_2=0.9$ and $w_1=0.5$, $w_2=0.5$.  It is observed that the algorithm converges under all cases. { While, the algorithm with adaptive initialization offers a higher social utility compared with uniform initialization and it has the result very close to the optimal randomized initialization.} This is because the adaptive initialization can potentially provide initial points that are closer to the optimum. Therefore, for all of the simulations in the following, adaptive initialization is used in the coordinate descent algorithm for the optimization in the scheme of full cooperation.

\subsubsection{Group Utility with  Partial Cooperation}
Next, we study the effects of selfish behaviours of an individual group on both itself and the other groups under partial cooperation. We examine two different kinds of behaviours of the groups in partial cooperation, namely, {\it selfish caching} and {\it un-selfish caching}. For the selfish group, its behaviour is defined in Section \ref{subsec:PartialCoop} in partial cooperation, which is to take advantage of the other groups with the knowledge of the social preference $R_f,~f\inm{F}$.  In contrast, for the un-selfish group, it faithfully caches files according to their request distribution (i.e., $\mv{c}_k=\mv{r}_k$). We consider the following utilities for evaluating the performances of group 1 and 2.
\begin{itemize}
    \item [1.]{\bf Group 1 utility under un-selfish caching}: $\cU_1(\mv{r}_1;\mv{r}_2)$;
    \item [2.]{\bf Group 2 utility under un-selfish caching}: $\cU_2(\mv{r}_2;\mv{r}_1)$;
    \item [3.]{\bf Group 1 utility under selfish group 1 caching}: $\cU_1(\mv{c}_1^*;\mv{r}_2)$;
    \item [4.]{\bf Group 2 utility under selfish group 1 caching}: $\cU_2(\mv{r}_2;\mv{c}_1^*)$.
\end{itemize}
\begin{figure}[t]
  \centering
  \includegraphics[width=9cm]{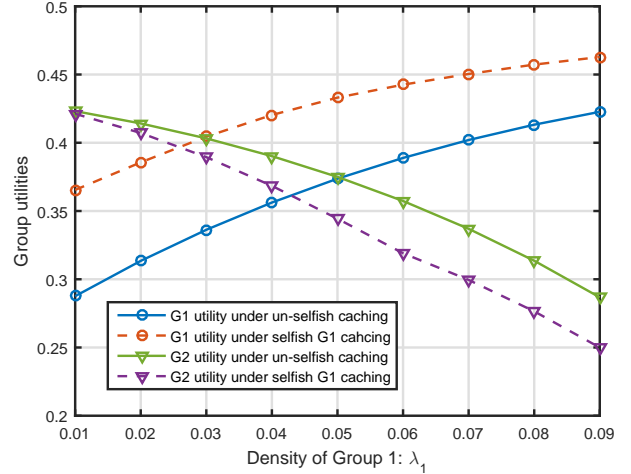}
  \caption{Group utilities in partial cooperation under different densities.}\label{fig1}
\end{figure}
The simulation result is shown in Fig. \ref{fig1} by varying the density $\lambda_1$ of group 1, while keeping the sum density of the two groups $\lambda_1+\lambda_2=\lambda_0$ fixed. It can be observed that the utilities of both groups monotonically increase with their own densities. This is obvious because higher group density leads to larger probability for file sharing within the group of MTs with the same file preference.  It can also be observed  that, compared with the utilities of  unselfish groups 1 and 2 ($\cU_1(\mv{r}_1;\mv{r}_2)$ and $\cU_2(\mv{r}_2;\mv{r}_1)$), the  utility $\cU_1(\mv{c}_1^*;\mv{r}_2)$ of selfish group 1 increases, while  the utility $\cU_2(\mv{r}_2;\mv{c}_1^*)$ of the un-selfish group 2 decreases. This shows that if a selfish group can exploit the social information $R_f$, it can indeed increase its own utility, while this may decrease the utility of the other un-selfish group. Hence, the simulation results in Fig. \ref{fig1} verify that there is indeed a conflict of interests between groups under heterogeneous file preferences. This motivates us to investigate the trade-offs between different groups, which is presented in the next simulation.

\subsubsection{Feasible Utility Region under Full Cooperation and the Benchmark Cases}
Next, we show the feasible utility region defined in (\ref{delayregion}) with Algorithm I.  The densities for the two groups are $\lambda_1=\lambda_2=0.05$ MT per square meter.  The Pareto boundary is obtained by solving optimization problems $\mathrm{(P1)}$ with varying weights $w_1$ and $w_2$ of the two groups. Specifically, the results for weights $w_1=1,~w_2=0$ and $w_1=0,~w_2=1$ are obtained based on  Theorem \ref{zeroTheorem} and the results for the weights $w_1,~w_2>0$ are obtained based on Theorem \ref{SocialTheorem}.  In comparison, the results for partial cooperation and no cooperation are also shown, which are based on Theorem \ref{theorem2} and Corollary \ref{CorollaryNonCoop}, respectively.

\begin{figure}[t]
  \centering
  \includegraphics[width=9cm]{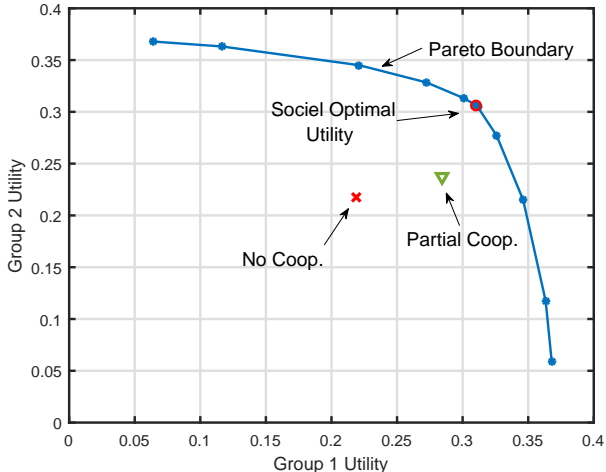}
  \caption{Group utility region under full cooperation versus utilities under  partial and no cooperation.}\label{paretoboundary}
\end{figure}
The simulation result for the feasible utility region is shown in Fig. \ref{paretoboundary}. It can be  observed that there is indeed a trade-off between the two groups in their utilities under full cooperation in cooperative local caching. Moreover, the group utilities in  partial cooperation and no cooperation both lie within the Pareto boundary achieved by full cooperation, which are at $(0.2839,0.2371)$ and $(0.2189,0.2173)$, respectively, both strictly lower than the social optimum utility at $(0.2937,0.2941)$ and lies inside the Pareto boundary.  This shows the benefits of full cooperation, while  partial  and no cooperation may lead to decreased utilities strictly inside the Pareto region. Finally, it should be noted that the utility of group 1 at $w_1=0$ or that of group 2 at $w_2=0$ is non-zero. This is because, the preferences of the two groups are not completely different and cooperative file sharing still can bring utilities to the group with even zero weight.

\subsubsection{Social Utility under Different Zipf Exponents}
Furthermore, we evaluate the social utility under the schemes of full, partial and no cooperation with respect to (a) different Zipf exponents and (b) different social densities. In case (a), since the typical Zipf exponent is around $\gamma_r=0.9$, we assess the influences of the Zipf exponent in the range of $0.2\leq \gamma_r\leq 1.8$. The densities for the two groups are $\lambda_1=\lambda_2=0.05$ MT per unit area.  In case (b), the densities of the two groups are $\lambda_1=\lambda_2=\lambda_0/2$ with varying $\lambda_0$ in the range of $0.02\leq \lambda_0\leq 0.18$.

\begin{figure*}[t]
\centering
\begin{subfigure}{.5\textwidth}
  \centering
  \includegraphics[width=8.5cm]{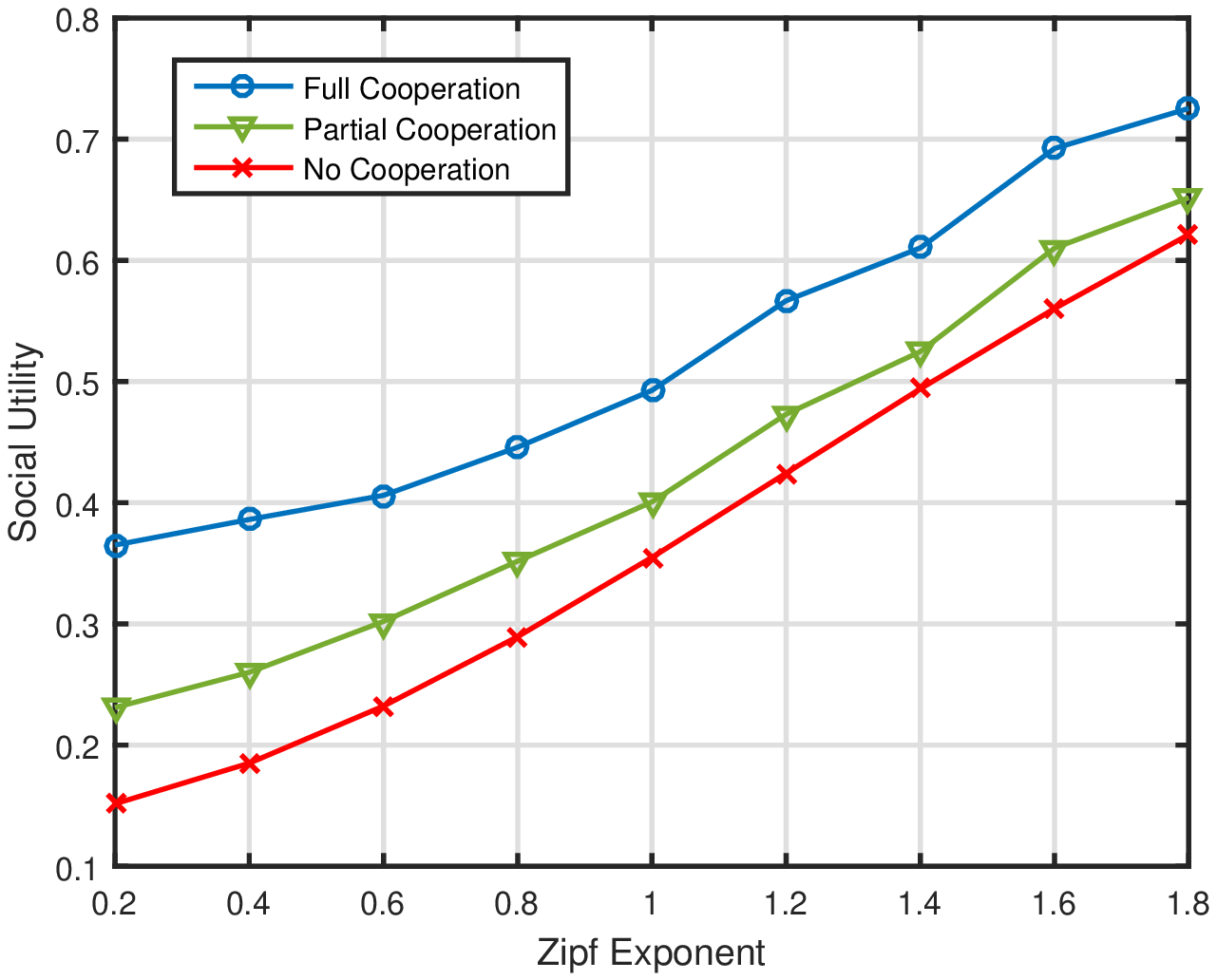}
  \caption{}\label{finalfigure1}
\end{subfigure}%
\begin{subfigure}{.5\textwidth}
  \centering
  \includegraphics[width=8.5cm]{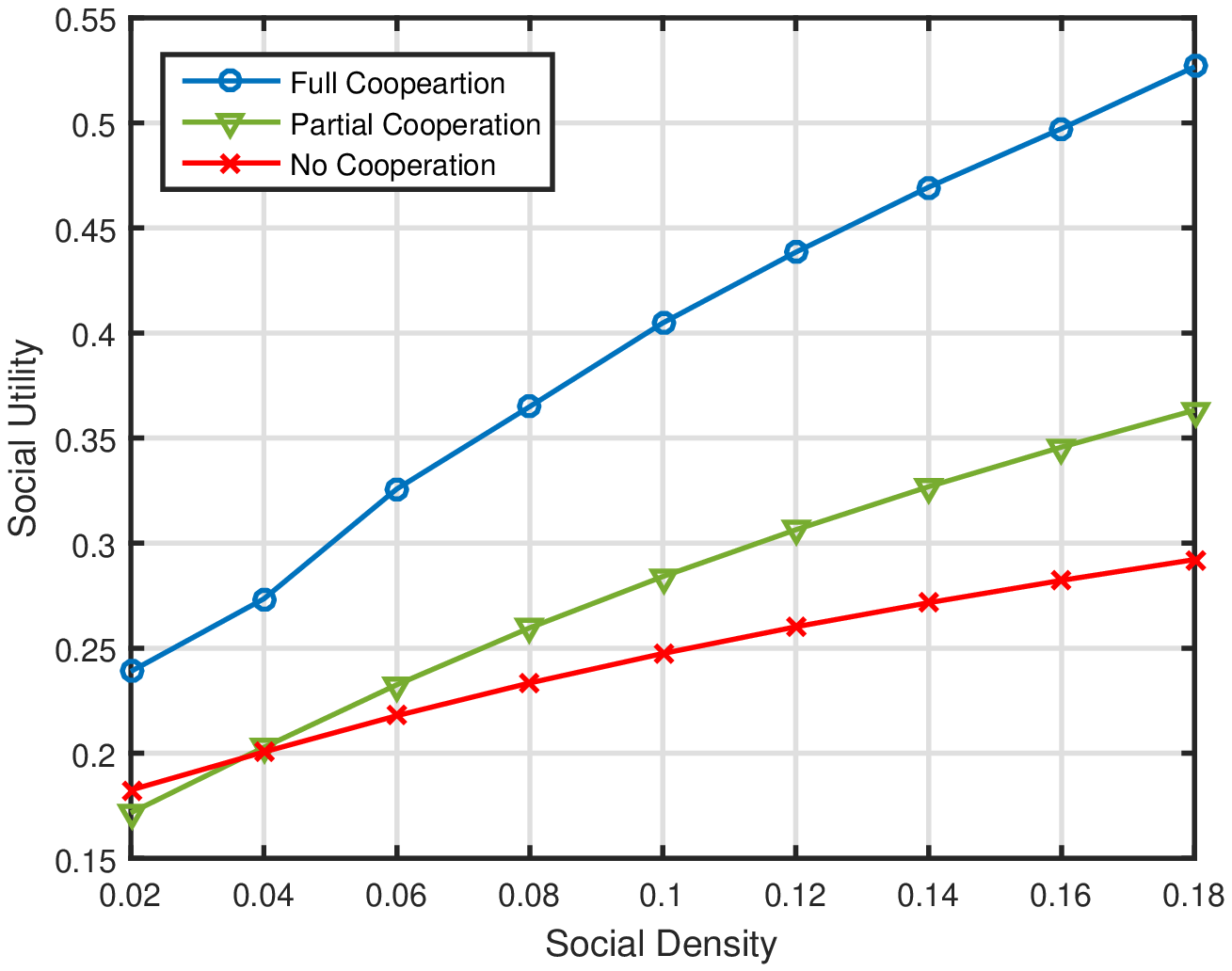}
  \caption{}\label{finalfigure2}
\end{subfigure}
\caption{Social utilities under (a) different Zipf exponents, and (b) different social densities.}\label{Finalfigure}
\end{figure*}
The simulation results are shown in Fig. \ref{Finalfigure}. First, from the result in Fig. \ref{finalfigure1}, it can be observed that for all of the three cases, the social utilities are  monotonically increasing with respect to the Zipf exponent. This can be explained by the majorization theory \cite{marshall2010inequalities}. With a large Zipf exponent, the popularity of the files is more concentrated on fewer files and the request distributions ($\mv{r}_1$ and $\mv{r}_2$) of the two groups will be more different.  Accordingly, both groups will try to match their caching distributions $\mv{c}_k$  with the request distributions $\mv{r}_k$. Hence, if a request distribution is more concentrated (i.e., it majorizes over another distribution that is less concentrated),  the utility will be lower. It can also be observed that the gap between the full cooperation and partial or no  cooperation decreases with the increase of the Zipf exponent (i.e., request distributions become more different). This is because, for partial cooperation and no cooperation, as the interests between different groups become more similar, the distribution on the popular files is more concentrated and it is easier for the MTs to discover these files. However, for the cases where the Zipf component is low, the popular files are more diverse and cooperation is more beneficial; thus, the gap between full cooperation and no or partial cooperation gets larger.

Next, from the result in Fig. \ref{finalfigure2}, it can be observed that, for all the three schemes, the social utility increases with respect to the social density. This is obvious since higher social density means larger number of MTs in the neighbour and more opportunities for file sharing and the gain is more substantial for the full cooperation. It can also be observed that partial cooperation can be very close or even worse than no cooperation in the low-density region. This is because in partial cooperation, groups try to exploit the distribution of the other groups, which may not be their actual caching distributions. In this case, the caching distribution may deviate far from the optimal caching distribution. As a result, the MTs' own caches are not matched to their group request distribution and they could get little help from the other groups. While for no cooperation, each group can at least cache according to its own preferences, which promotes intra-group file sharing.

\subsection{Network Simulation}
In this sub-section, we perform a network-level simulation under the evolution of the system to validate the accuracy of the theoretical results for the cooperative local caching under three cases. The simulation set-up is specified as follows: three groups, each having 100 MTs, 300 MTs, 600 MTs, are scattered within a square area of $100\times100~\mathrm{m}^2$. Hence, the density of the three groups are $\lambda_1=0.01,\lambda_2=0.03,\lambda_3=0.06$ MTs per square meter and their weights are $\mv{w}=[0.1,0.3,0.6]$, respectively. Initially, the locations of all the MTs are randomly generated within this area. The simulation is operated under 300 time slots. {During each time slot,  each MT randomly chooses a direction from $[0,2\pi]$ and moves $1~\mathrm{m}$ in that direction.} We first obtain the optimal caching distribution $c_{k,f}^*,~f\inm{F}$ for each group $k\inm{K}$ in full, partial and no cooperation with the results from Theorem \ref{SysModel:delay}, Theorem \ref{theorem2} and Corollary \ref{CorollaryNonCoop}, respectively. According to these optimal distributions, which are multinomial distributions with one trail, the MTs in each group perform random samplings on the set of files for obtaining the cached file and they keep the cached file during the whole simulation. The file request during each time slot is also a random sampling from the request distribution  $r_{k,f},~f\inm{F}$ of each group $k\inm{K}$. The MTs search for the requested file within the range of $d=5\mathrm{m}$. During time slot $n$, utility of each group $k\inm{K}$ is defined as

\begin{figure*}[t]
\centering
\begin{subfigure}{.33\textwidth}
  \centering
  \includegraphics[width=5.9cm]{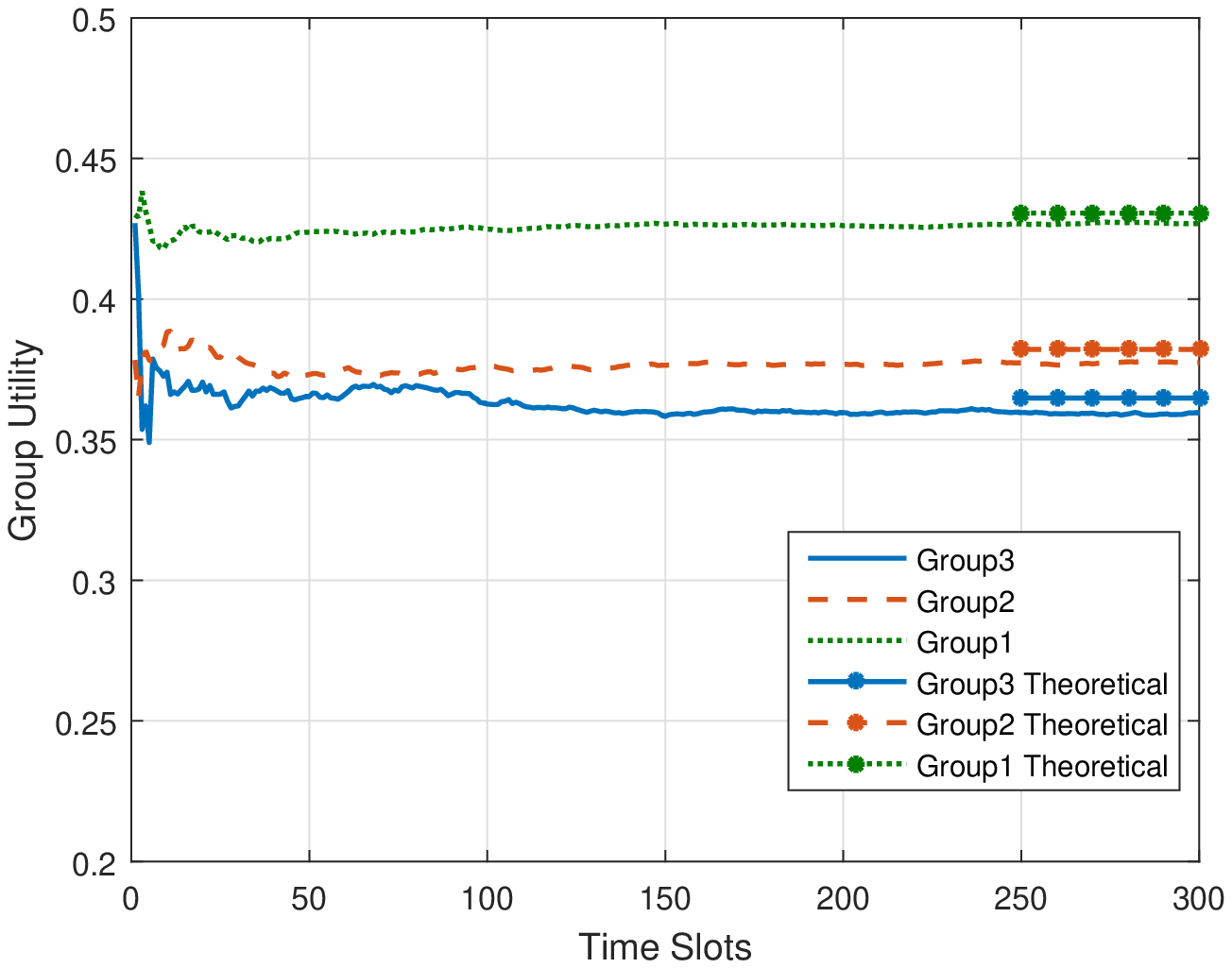}
  \caption{}\label{networksim1}
\end{subfigure}%
\begin{subfigure}{.33\textwidth}
  \centering
  \includegraphics[width=5.9cm]{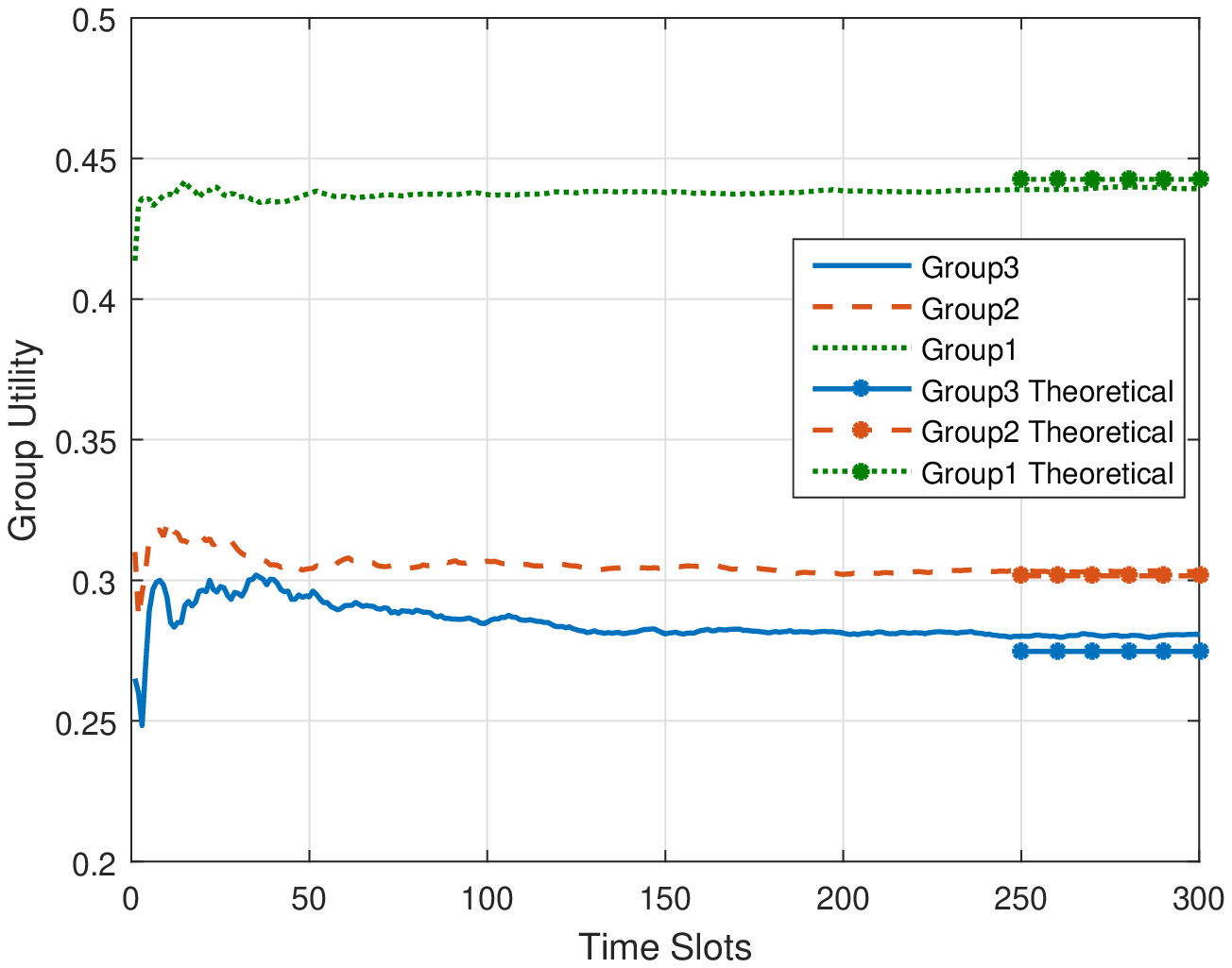}
  \caption{}\label{networksim2}
\end{subfigure}
\begin{subfigure}{.33\textwidth}
  \centering
  \includegraphics[width=5.9cm]{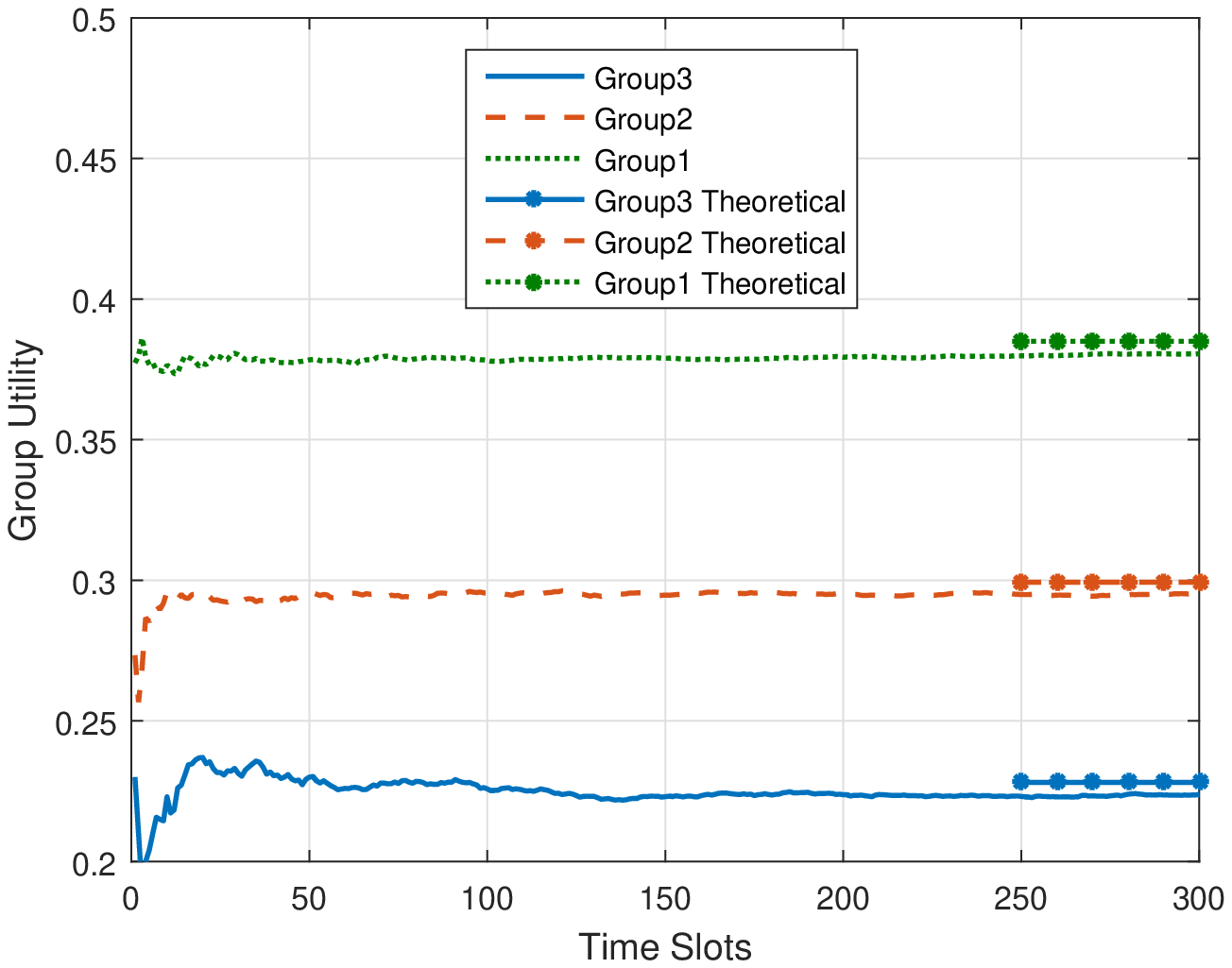}
  \caption{}\label{networksim3}
\end{subfigure}
\caption{Group utilities under (a) full cooperation, (b) partial cooperation, and (c) no cooperation.}\label{networksim}
\end{figure*}
\begin{figure*}
\begin{align}
 \cU_k(n)=\frac{\text{Total successful file discoveries of group $k$ up to time slot $n$}}{\text{Total file requests of group $k$ up to time slot $n$}}.
\end{align}
\hrule
\end{figure*}
With results in Theorem \ref{SysModel:delay}, Theorem \ref{theorem2} and Corollary \ref{CorollaryNonCoop},  three groups' utilities under full, partial and no cooperation are  obtained as $(0.3648, 0.3821, 0.4306)$, $(0.2747,0.3016,0.4426)$ and $(0.2281,0.2993,0.3850)$, respectively. The network simulation results after 300 time slots for full cooperation, partial cooperation and no cooperation are shown in Fig. \ref{networksim1}, \ref{networksim2} and \ref{networksim3}, respectively. From the simulation results, it can be observed that the converged values of the cooperative local caching system under real-time simulation are very close to their corresponding theoretical results. This validates the effectiveness of our system under real operation and shows the accuracy of our model.

}
\section{Conclusions and Future Work}\label{sec:conclusion}
Our studies in this paper are motivated by the lack of investigation in the literature on the cooperative local caching with heterogeneous file preferences among different MTs. We practically categorize the MTs into different groups according to their preferences and investigate the trade-offs between different groups. We first consider the case of full cooperation, which is formulated as a weighted-sum utility maximization problem to characterize the optimal trade-offs between different groups. We also study two benchmark cases under selfish caching: namely, partial and no cooperation with and without inter-group file sharing, respectively.  Closed-form solutions are obtained for these two benchmark cases by utilizing the KKT conditions. Finally, extensive numerical examples are presented to compare the three cases and show the effectiveness of the fully cooperative local caching compared to the benchmark cases. In summary, our work provides essential insights in understanding the trade-offs among different groups and the effects of their selfish behaviours under heterogeneous file preferences.  The design insights provided apply to networks under a general setting, including but not limited to wireless cellular networks.

{Despite the new insights provided by this work, there are still some possible directions for the future work. First, in this work, we consider the cooperative local caching between MTs within one hop under heterogeneous file preference. It would be interesting to generalize the results to multiple hops, which will further foster the cooperation between MTs. Second, in this work we consider unicast of the files sharing under heterogeneous file preference. It will be interesting yet challenging to extend to the case of multicast file sharing \cite{JiCM14DundLimitD2D}.  Last but not the least, considering that different groups have conflicts of interests under heterogeneous file preference, it is also pertinent to consider the behaviour of the MTs in cooperative caching from the perspective of {\it game theory}.
}
\appendices
{\section{Proof for Theorem \ref{SysModel:delay}}\label{Appendix:proofthm1}
Consider the typical MT  $i\in\cG_k,~k\inm{K}$ in group $k$. We denote the MTs from group $j\inm{K}$ that falls into the range of MT $i$ in group $k$ as $\cN_j=\{l\inm{K}_j|X_{j,l}\in B(X_{k,i},d),~j\inm{K}\}\subseteq \cG_j$ and denote the number of MTs in this set as $N_j=|\cN_j|$.  Please be noted that because of the Slyvnyak-Mecke Theorem \cite{baccelli2009stochastic}, the distribution of the homogeneous Poisson point process $\Phi_k$ of group $k$ conditioning on any particular point, is the same as the distribution of the point process $\Phi_k$. Hence, although we are conditioning on the typical MT $i\inm{G}_k$ for file request in group $k$, the density of the point process $\Phi_k$ is still $\lambda_k$.  According to the model of HPPP, it follows that $N_j$ is a Poisson random variable with mean $\mu_k=\pi d^2\lambda_k,~k\inm{K}$ and its PMF is given by
\begin{align}
  \mathbb{P}[N_j]=\frac{\mu_{j}^{N_j}}{N_j!}e^{-\mu_{j}},~N_j=0,1,\cdots,\infty.\nonumber
\end{align}
With the theorem of iterated expectation, group $k$'s utility  is
\begin{align}
\cU_k(\mv{c}_k;\mv{c}_{-k})=\sum_{f\inm{F}}r_{k,f}\mP(E|f;\mv{c}_k,\mv{c}_{-k}).
\end{align}
Then, given the requested file $f\inm{F}$ and the neighbouring MTs $\cN_j\subseteq\cG_j,~j\inm{K}$, the probability for successful file discovery is
\begin{align}
&\mP(E|f,\{\cN_j\};\mv{c}_k,\mv{c}_{-k})=\cacheprob+(1-{\cacheprob})\Big[1-\prod_{j\inm{K}}(1-{c_{j,f}})^{N_j}\Big]\nonumber\\
&=1- \bar{c}_{k,f}\prod_{j\inm{K}}\bar{c}_{j,f}^{N_j},\nonumber
\end{align}
where $\prod_{k\inm{K}}(1-{c_{k,f}})^{N_k}$ denotes the probability that file $f\inm{F}$ is not found in caches of the neighbouring MTs from all groups.   Then, the probability of successful file discovery is
\begin{align}
&\mP(E;\mv{c}_k,\mv{c}_{-k})=\sum_{f\inm{F}}r_{k,f}\mP(E|f;\mv{c}_k,\mv{c}_{-k})\nonumber\\
&=\sum_{f\inm{F}}r_{k,f}\sum_{N_1}\mathbb{P}[N_1]\sum_{N_2}\mathbb{P}[N_2]\cdots\sum_{N_K}\mathbb{P}[N_K]\nonumber\\
&\times \mathbb{P}[E|f,\{\cN_j\};\mv{c}_k,\mv{c}_{-k}]\nonumber\\
&=\sum_{f\inm{F}}r_{k,f}\sum_{N_1=0}^\infty e^{-\mu_{1}}\frac{\mu_{1}^{N_1}}{N_1!}\sum_{N_2=0}^\infty e^{-\mu_{2}}\frac{\mu_{2}^{N_2}}{N_2!}\cdots\sum_{N_K=0}^\infty e^{-\mu_{K}}\frac{\mu_{K}^{N_K}}{N_K!}\nonumber\\
&\times\Big(1- \bar{c}_{k,f}\prod_{j\inm{K}}\bar{c}_{j,f}^{N_j}\Big)\nonumber\\
&\stackrel{(a)}{=}\sum_{f\inm{F}}\reqprob\left(1-\bar{c}_{k,f}e^{-\mu_0C_{f}}\right),\nonumber
\end{align}
where  ${(a)}$ is obtained by utilizing (\ref{SocialCaching}) and the Taylor expansion for exponential function $e^{x}=\sum_{i=0}^{\infty}{x^i}/{i!}$. Theorem \ref{SysModel:delay} is thus proved.}

\section{Proof for Theorem \ref{SocialTheorem}}\label{sec:proofthm2}
First, it should be noted that maximizing the objective function in the original problem is equivalent to minimizing the objective function $\sum_{j\inm{K}_+}w_j\sum_{f\inm{F}}r_{j,f}\bar{c}_{j,f}e^{-\mu_0C_{f}}$. By introducing the dual variable $\gamma_k$ for the constraint in (\ref{p2robust}), the partial Lagrangian for problem  $\mathrm{(P2-1)}$ is
\begin{align}
&\cL\left(\mv{c}_k,\gamma_k\right)=\sum_{j\inm{K}_+}w_j\sum_{f\inm{F}}r_{j,f} \bar{c}_{j,f}e^{-\mu_0}e^{\sum_{k\inm{K}}\mu_k\bar{c}_{k,f}}\nonumber\\
&-\gamma_k\Big(\sum_{f\inm{F}} {\bar{c}_{k,f}}-F+1\Big)\nonumber\\
&=\sum_{f\inm{F}}\Bigg(\sum_{j\inm{K}_+}w_jr_{j,f} \bar{c}_{j,f}e^{-\mu_0}e^{\sum_{k\inm{K}}\mu_k\bar{c}_{k,f}}-\gamma_k {\bar{c}_{k,f}} \Bigg)\nonumber\\
&+\gamma_k(F-1).
\end{align}
 The dual function is then given as
\begin{align}
g(\gamma_k)=\mathop{\mathtt{min.}}_{\mv{\bar c}_k}&~~\cL\left(\mv{c}_k,\gamma_k\right)\nonumber\\
\mathtt{s.t.}&~~0\leq \bar{c}_{k,f}\leq 1,~ f\inm{F}.
\end{align}
The associated dual problem is then defined as
\begin{align}
\mathop{\mathtt{max.}}_{\gamma_k\geq 0}~~g(\gamma_k).
\end{align}
Because the primal problem is convex and satisfies the Slater's condition \cite{boyd2004convex}, the duality gap between the primal and dual problem is zero. Hence, the problem can be optimally solved in the dual domain by first minimizing the Lagrangian $\cL\left(\mv{c}_k,\gamma_k\right)$ for a given $\gamma_k$ and then maximizing the dual function $g(\gamma_k)$ with respect to $\gamma_k$. For a given $\gamma_k$, the dual function can be decomposed into parallel sub-problems, each for a given file $f\inm{F}$ as follows.
\begin{align}
\mathrm{(P2-2)}:\nonumber\\
\mathop{\mathtt{min.}}_{0\leq \cacheprobcomp\leq 1}&~~ \sum_{j\inm{K}_+}w_jr_{j,f}\bar{c}_{j,f}e^{-\mu_0}e^{\sum_{k\inm{K}}\mu_k\bar{c}_{k,f}}-\gamma_k {\bar{c}_{k,f}}
\end{align}

First,  according to the KKT conditions, the following equations should be satisfied by the optimal primal and dual solutions,
\begin{align}
\gamma_k^*\left(\sum_{f\inm{F}} {\bar{c}_{k,f}^{*}}-F+1\right)&=0,\label{1kkt1}\\
0\leq {\bar{c}_{k,f}^{*}}&\leq 1,~f\inm{F}.\label{1kkt2}
\end{align}
Then,  the Lagrangian for the sub-problem of file $f\inm{F}$ is
\begin{align}
&\cL_{f}({\bar{c}_{k,f}},\nu_{k,f,l},\nu_{k,f,r})=\sum_{j\inm{K}_+}w_jr_{j,f}{\bar{c}_{j,f}}e^{-\mu_0}e^{\sum_{k\inm{K}}\mu_k\bar{c}_{k,f}}\nonumber\\
&-\gamma_k {\bar{c}_{k,f}}-\nu_{k,f,l}{\bar{c}_{k,f}}+\nu_{k,f,r}({\bar{c}_{k,f}}-1),
\end{align}
where $\nu_{k,f,l}\geq 0$ is the dual variable for  ${\cacheprobcomp}\geq 0$, and $\nu_{k,f,r}\geq 0$ is the dual variable for ${\cacheprobcomp}\leq 1$. By taking the partial derivative of the Lagrangian with respect to ${\cacheprobcomp}$, we obtain
\begin{align}\label{lagrangian}
&\frac{\partial\mathcal{L}_{f}}{\partial {\cacheprobcomp}}=e^{-\mu_0}e^{\sum_{j\inm{K}}\mu_j\bar{c}_{j,f}} \Bigg(w_kr_{k,f}+\nonumber\\
&\sum_{j\inm{K}_+\setminus\{k\}}w_jr_{j,f}\bar{c}_{j,f}\mu_k+w_kr_{k,f}\mu_k\bar{c}_{k,f} \Bigg)-\gamma_k-\nu_{k,f,l}+\nu_{k,f,r}.
\end{align}
By the KKT conditions, we can obtain the following system of equations:
\begin{align}
\frac{\partial\mathcal{L}_{f}}{\partial {\cacheprobcomp}}&=0,\label{2kkt1}\\
\nu_{k,f,l}^*{\bar{c}_{k,f}^{*}}=0,~\nu_{k,f,r}^*({\bar{c}_{k,f}^{*}}-1)&=0,\label{2kkt2}\\
\nu_{k,f,l}^*,\nu_{k,f,r}^*&\geq 0.\label{2kkt3}
\end{align}
With (\ref{2kkt1}), we can obtain that
\begin{align}
e^{-\mu_k\bar{c}_{k,f}}=\frac{A'_{k,f}}{C'_{k,f}}+\frac{B'_{k,f}}{C'_{k,f}}\bar{c}_{k,f},
\end{align}
where
\begin{align}
A'_{k,f}&=\left(w_kr_{k,f}+\sum_{j\inm{K}_+\setminus\{k\}}w_jr_{j,f}\bar{c}_{j,f}\mu_k\right)\nonumber\\
&\times e^{-\mu_0}e^{\sum_{j\inm{K}\setminus\{k\}}\mu_j\bar{c}_{j,f}}\\
B'_{k,f}&=w_kr_{k,f}\mu_{k}e^{-\mu_0}e^{\sum_{j\inm{K}\setminus\{k\}}\mu_j\bar{c}_{j,f}},\\
C'_{k,f}&=\gamma_k+\nu_{k,f,l}-\nu_{k,f,r}.
\end{align}
By utilizing the equation $e^{ax+b}=cx+d\rightarrow x=-\frac{1}{a}\cW(-\frac{a}{c}e^{b-\frac{ad}{c}})-\frac{d}{c}$, where $\cW(\cdot)$ is the Lambert-W function \cite{corless1996lambertw}, $a=-\mu_k$, $b=0$, $c=\frac{B'_{k,f}}{C'_{k,f}}$ and $d=\frac{A'_{k,f}}{C'_{k,f}}$. Hence, it follows that
\begin{align}
\bar{c}_{k,f}=\frac{1}{\mu_k}\cW\Big(A''_{k,f}e^{\mu_kB_{k,f}}\Big)-{B_{k,f}},
\end{align}
where
\begin{align}
A''_{k,f}&=\frac{\mu_kC'_{k,f}}{B'_{k,f}}=\frac{\gamma_k+\nu_{k,f,l}-\nu_{k,f,r}}{w_ke^{-\mu_0}r_{k,f}e^{\sum_{j\inm{K}\setminus\{k\}}\mu_j\bar{c}_{j,f}}},\\
B_{k,f}&=\frac{A'_{k,f}}{B'_{k,f}}=\frac{1}{\mu_k}+\sum_{j\inm{K}_+\setminus\{k\}}\frac{w_jr_{j,f}}{w_kr_{k,f}}\bar{c}_{j,f}.\label{resultb}
\end{align}
 For notational convenience, in the following, we denote
\begin{align}
&{\tilde{c}_{k,f}}(\gamma_k ,\nu_{k,f,l},\nu_{k,f,r})=\frac{1}{\mu_k}\cW\Big(A''_{k,f}e^{\mu_kB_{k,f}}\Big)-{B_{k,f}}.
\end{align}
Because the Lambert-W function is monotonically increasing, the function ${\tilde{c}_{k,f}}(\gamma_k,\nu_{k,f,l},\nu_{k,f,r})$ is monotonically increasing with respect to $\nu_{k,f,l}$ and decreasing with respect to $\nu_{k,f,r}$.

Then, we discuss the following three cases of ${\tilde{c}_{k,f}}(\gamma_k,\nu_{k,f,l},\nu_{k,f,r})$  on the regions of $(-\infty,0]$, $(0,1)$ and $[1,\infty)$, respectively.
\begin{itemize}
\item First, for the case of $(-\infty,0]$, assume that $ {\bar{c}_{k,f}^{*}}>0$ when  ${\tilde{c}_{k,f}}(\gamma_k,0,0)\leq 0$. When $ {\bar{c}_{k,f}^{*}}>0$, $\nu_{k,f,l}^*=0$ should be satisfied due to the complementary slackness condition in (\ref{2kkt2}). Hence, ${\tilde{c}_{k,f}}(\gamma_k,0,\nu_{k,f,r})>0$. Moreover, because ${\tilde{c}_{k,f}}(\gamma_k,0,\nu_{k,f,r}^*)$ is monotonically decreasing with respect to $\nu_{k,f,r}$ and $\nu_{k,f,r}\geq 0$, we obtain ${\tilde{c}_{k,f}}(\gamma_k,0,0)\geq {\tilde{c}_{k,f}}(\gamma_k,0,\nu_{k,f,r}^*)>0$. This contradicts the previous assumption. Moreover, due to the primal feasibility $ {\bar{c}_{k,f}^{*}}\geq 0$ in (\ref{1kkt2}), when  ${\tilde{c}_{k,f}}(\gamma_k,0,0)\leq 0$, $ {\bar{c}_{k,f}^{*}}=0$ is the optimal solution;
\item  Next, for the case of $(0,1)$,  $\nu_{k,f,r}=\nu_{k,f,l}=0,~f\inm{F}$ due to the complementary slackness condition in (\ref{2kkt2}). Hence, $ {\bar{c}_{k,f}^{*}}=\tilde{c}_{k,f}(\gamma_k,0,0)$;
\item Finally, for the case of $[1,\infty)$, similar to the first case, it can be proved by contradiction that, when  ${\tilde{c}_{k,f}}(\gamma_k,0,0)\geq 1$, $ {\bar{c}_{k,f}^{*}}=1$.
\end{itemize}
By summarizing the above three cases, we can obtain that the optimal file caching probability for file $f\inm{F}$ of group $\cG_k,~k\inm{K}$ is
\begin{align}\label{appen:result1}
{c_{k,f}^{*}}&=\left[1-{{\tilde{c}_{k,f}}(\gamma_k,0,0)}\right]^{1}\nonumber\\
&=\left[1-\Bigg(\frac{1}{\mu_k}\cW\Big(A_{k,f}e^{\mu_kB_{k,f}}\Big)-{B_{k,f}}\Bigg)\right]_0^1,
\end{align}
 where
 \begin{align}
 A_{k,f}=\frac{\gamma_k}{w_ke^{-\mu_0}r_{k,f}e^{\sum_{j\inm{K}\setminus\{k\}}\mu_j\bar{c}_{j,f}}}
 \end{align}
 and $B_{k,f}$ is defined in (\ref{resultb}). With the above results, the optimal dual variable $\gamma^*_k$ can be obtained by the constraint $\sum_{f\inm{F}}{c}_{k,f}^*=1$. Theorem \ref{SocialTheorem} is thus proved.

\def\UrlBreaks{\do\/\do-}
\bibliographystyle{IEEEtran}
\bibliography{Yinghao4}

\begin{thebibliography}{10}
\providecommand{\url}[1]{#1}
\csname url@samestyle\endcsname
\providecommand{\newblock}{\relax}
\providecommand{\bibinfo}[2]{#2}
\providecommand{\BIBentrySTDinterwordspacing}{\spaceskip=0pt\relax}
\providecommand{\BIBentryALTinterwordstretchfactor}{4}
\providecommand{\BIBentryALTinterwordspacing}{\spaceskip=\fontdimen2\font plus
\BIBentryALTinterwordstretchfactor\fontdimen3\font minus
  \fontdimen4\font\relax}
\providecommand{\BIBforeignlanguage}[2]{{%
\expandafter\ifx\csname l@#1\endcsname\relax
\typeout{** WARNING: IEEEtran.bst: No hyphenation pattern has been}%
\typeout{** loaded for the language `#1'. Using the pattern for}%
\typeout{** the default language instead.}%
\else
\language=\csname l@#1\endcsname
\fi
#2}}
\providecommand{\BIBdecl}{\relax}
\BIBdecl

\bibitem{CISCOVNI}
\BIBentryALTinterwordspacing
Cisco, ``Cisco visual networking index: Global mobile data traffic forecast
  update 2014–-2019 white paper,'' 2015. [Online]. Available:
  \url{http://www.cisco.com/c/en/us/solutions/collateral/service-provider/visual-networking-index-vni/white\_paper\_c11-520862.pdf}
\BIBentrySTDinterwordspacing

\bibitem{informa2008mobile}
{Informa (Firm). Informa Telecoms \& Media}, \emph{Mobile Broadband Access at
  Home: The Business Case for {Femtocells, UMA and IMS-VCC Dual Mode
  Solutions}}.\hskip 1em plus 0.5em minus 0.4em\relax Informa Telecoms \&
  Media, 2008.

\bibitem{LocalCachingComMag}
N.~Golrezaei, A.~Molisch, A.~Dimakis, and G.~Caire, ``Femto-caching and
  device-to-device collaboration: {a} new architecture for wireless video
  distribution,'' \emph{IEEE Commun. Mag.}, vol.~51, no.~4, pp. 142--149, Apr.
  2013.

\bibitem{Dowdy}
L.~W. Dowdy and D.~V. Foster, ``Comparative models of the file assignment
  problem,'' \emph{ACM Comput. Surv.}, vol.~14, no.~2, pp. 287--313, Jun. 1982.

\bibitem{FundLimitofCaching}
M.~Maddah-Ali and U.~Niesen, ``Fundamental limits of caching,'' \emph{IEEE
  Trans. Inf. Theory}, vol.~60, no.~5, pp. 2856--2867, May 2014.

\bibitem{TITFemto}
K.~Shanmugam, N.~Golrezaei, A.~Dimakis, A.~Molisch, and G.~Caire,
  ``{Femto-Caching}: {wireless} content delivery through distributed caching
  helpers,'' \emph{IEEE Trans. Inf. Theory}, vol.~59, no.~12, pp. 8402--8413,
  Dec. 2013.

\bibitem{AdaptiveVIdeoStreaming}
D.~Bethanabhotla, G.~Caire, and M.~Neely, ``Adaptive video streaming for
  wireless networks with multiple users and helpers,'' \emph{IEEE Trans.
  Commun.}, vol.~63, no.~1, pp. 268--285, Jan. 2015.

\bibitem{TaoMeixiaYuWei}
M.~Tao, E.~Chen, H.~Zhou, and W.~Yu, ``Content-centric sparse multicast
  beamforming for cache-enabled cloud {RAN},'' \emph{IEEE Trans. Wireless
  Commun.}, vol.~15, no.~9, pp. 6118--6131, Sep. 2016.

\bibitem{TITD2Dcaching}
N.~Golrezaei, A.~Dimakis, and A.~Molisch, ``Scaling behavior for
  device-to-device communications with distributed caching,'' \emph{IEEE Trans.
  Inf. Theory}, vol.~60, no.~7, pp. 4286--4298, Jul. 2014.

\bibitem{JiCM13}
M.~Ji, G.~Caire, and A.~F. Molisch, ``The throughput-outage tradeoff of
  wireless one-hop caching networks,'' \emph{IEEE Trans. Inf. Theory}, vol.~61,
  no.~12, pp. 6833--6859, Dec. 2015.

\bibitem{JiCM14DundLimitD2D}
------, ``Fundamental limits of caching in wireless {D2D} networks,''
  \emph{IEEE Trans. Inf. Theory}, vol.~62, no.~2, pp. 849--869, Feb. 2016.

\bibitem{Multihop}
S.~Gitzenis, G.~Paschos, and L.~Tassiulas, ``Asymptotic laws for joint content
  replication and delivery in wireless networks,'' \emph{IEEE Trans. Inf.
  Theory}, vol.~59, no.~5, pp. 2760--2776, May 2013.

\bibitem{BoHanPanHui}
B.~Han, P.~Hui, V.~Kumar, M.~Marathe, J.~Shao, and A.~Srinivasan, ``Mobile data
  offloading through opportunistic communications and social participation,''
  \emph{IEEE Trans. Mob. Comput.}, vol.~11, no.~5, pp. 821--834, May 2012.

\bibitem{ChenProulx}
X.~Chen, B.~Proulx, X.~Gong, and J.~Zhang, ``Exploiting social ties for
  cooperative {D2D} communications: {a} mobile social networking case,''
  \emph{IEEE/ACM Trans. Networking}, vol.~23, no.~5, pp. 1471--1484, Oct. 2015.

\bibitem{SocialCellular}
J.~Hu, L.~L. Yang, and L.~Hanzo, ``Delay analysis of social group
  multicast-aided content dissemination in cellular system,'' \emph{IEEE Trans.
  Commun.}, vol.~64, no.~4, pp. 1660--1673, Apr. 2016.

\bibitem{SocialVTC}
------, ``Distributed multistage cooperative-social-multicast-aided content
  dissemination in random mobile networks,'' \emph{IEEE Trans. Veh. Tech.},
  vol.~64, no.~7, pp. 3075--3089, Jul. 2015.

\bibitem{SocialD2D}
Y.~Li, T.~Wu, P.~Hui, D.~Jin, and S.~Chen, ``Social-aware {D2D} communications:
  qualitative insights and quantitative analysis,'' \emph{IEEE Commun. Mag.},
  vol.~52, no.~6, pp. 150--158, Jun. 2014.

\bibitem{JiConf}
M.~Ji, K.~Shanmugam, G.~Vettigli, J.~Llorca, A.~M. Tulino, and G.~Caire, ``An
  efficient multiple-groupcast coded multicasting scheme for finite fractional
  caching,'' in \emph{Proc. IEEE Int. Conf. Commun. (ICC)}, Jun. 2015, pp.
  3801--3806.

\bibitem{ChenLZT16}
Z.~Chen, Y.~Liu, B.~Zhou, and M.~Tao, ``Caching incentive design in wireless
  {D2D} networks: A stackelberg game approach,'' in \emph{Proc. IEEE Int. Conf.
  Commun. (ICC)}, May 2016, pp. 1--6.

\bibitem{bluetooth}
\BIBentryALTinterwordspacing
{Bluetooth SIG Inc.}, ``Bluetooth low energy, bluetooth development portal,''
  2014. [Online]. Available:
  \url{https://developer.bluetooth.org/TechnologyOverview/Pages/BLE.aspx}
\BIBentrySTDinterwordspacing

\bibitem{Kingman1993}
J.~Kingman, \emph{Poisson Process}.\hskip 1em plus 0.5em minus 0.4em\relax
  Oxford, United Kingdom: Oxford University Press, 1993.

\bibitem{LocalCachingCommMag}
X.~Wang, M.~Chen, T.~Taleb, A.~Ksentini, and V.~Leung, ``Cache in the air:
  exploiting content caching and delivery techniques for {5G} systems,''
  \emph{IEEE Commun. Mag.}, vol.~52, no.~2, pp. 131--139, Feb. 2014.

\bibitem{boyd2004convex}
S.~Boyd and L.~Vandenberghe, \emph{Convex Optimization}.\hskip 1em plus 0.5em
  minus 0.4em\relax Cambridge, United Kingdom: Cambridge University Press,
  2004.

\bibitem{corless1996lambertw}
R.~M. Corless, G.~H. Gonnet, D.~E. Hare, D.~J. Jeffrey, and D.~E. Knuth, ``On
  the {Lambert-W} function,'' \emph{Advances in Computational mathematics},
  vol.~5, no.~1, pp. 329--359, 1996.

\bibitem{goldsmith2005wireless}
A.~Goldsmith, \emph{Wireless communications}.\hskip 1em plus 0.5em minus
  0.4em\relax Cambridge, United Kingdom: Cambridge University Press, 2005.

\bibitem{ZipfDistribution}
L.~Breslau, P.~Cao, L.~Fan, G.~Phillips, and S.~Shenker, ``Web caching and
  {Zipf}-like distributions: evidence and implications,'' in \emph{Proc. 18th
  Annual Joint Conf. IEEE Comput. and Commun. Societies (INFOCOM '99)}, vol.~1,
  Mar. 1999, pp. 126--134.

\bibitem{marshall2010inequalities}
A.~W. Marshall, I.~Olkin, and B.~Arnold, \emph{Inequalities: theory of
  majorization and its applications}.\hskip 1em plus 0.5em minus 0.4em\relax
  Berlin, Germany: Springer Science \& Business Media, 2010.

\bibitem{baccelli2009stochastic}
F.~Baccelli and B.~Blaszczyszyn, \emph{Stochastic geometry and wireless
  networks: Vol. I Theory}.\hskip 1em plus 0.5em minus 0.4em\relax Delft,
  Netherlands: Now Publishers Inc., 2009.

\end{thebibliography}

\end{document}